%% file: curv_est_omts.tex
\documentclass[a4paper,fleqn,12pt,twoside,draft]{article}

\input{layout_commands}
\input{math_commands}

\input{my_commands}

\title{\titlefamily\Huge Curvature estimates for stable marginally trapped surfaces}
\author{\titlefamily Lars Andersson
  \thanks{Supported in part by the NSF, under contract no. DMS 0407732 with the University of Miami.}\\
  \titlefamily\small lars.andersson@aei.mpg.de\\
  \titlefamily\small Albert-Einstein-Institut,
  \titlefamily\small Am M\"uhlenberg 1, 14476 Potsdam, Germany.\\[1ex]
  \titlefamily\small Department of Mathematics, University of Miami,
  \titlefamily\small Coral Gables, FL 33124, USA.\\  
  \and
  \titlefamily Jan Metzger\\
  \titlefamily\small jan.metzger@aei.mpg.de\\
  \titlefamily\small Albert-Einstein-Institut, 
  \titlefamily\small Am M\"uhlenberg 1, D-14476 Potsdam, Germany.
}
\date{}

\begin{document}
\hyphenation{}
\pagestyle{footnumber}
\maketitle
\thispagestyle{footnumber}
\begin{abst}%
  We derive local integral and $\sup$- estimates for the curvature of
  stably marginally outer trapped surfaces in a sliced space-time. The
  estimates bound the shear of a marginally outer trapped surface in
  terms of the intrinsic and extrinsic curvature of a slice containing
  the surface. These estimates are well adapted to situations of
  physical interest, such as dynamical horizons.
\end{abst}
\input{intro}
%
\input{prelim}

%
\input{simons}
%
\input{linearization}

%
\input{stable}

%
\input{apriori}

%
\input{areabound}
%
\input{applic}

%
\input{ack}

%
\bibliographystyle{amsalpha}
\bibliography{references}
\end{document}

%% file: layout_commands.tex
\usepackage{amsmath}
\usepackage{amsthm}
\usepackage[shiftmargins]{vmargin}

\newcommand{\titlefamily}{\sffamily}
\newcommand{\thmfamily}{\sffamily}
\newcommand{\sectitlefnt}{\titlefamily\bfseries\Large}
\newcommand{\subsectitlefnt}{\titlefamily\bfseries}

\setpapersize{A4}
\setmarginsrb{35mm}{25mm}{30mm}{25mm}{0mm}{0mm}{10mm}{3\baselineskip}

\makeatletter
\renewcommand{\section}{\@startsection
  {section}%
  {1}%
  {0mm}%
  {-\baselineskip}%
  {\baselineskip}%
  {\sectitlefnt}}%
\renewcommand{\subsection}{\@startsection
  {subsection}%
  {2}%
  {0mm}%
  {-\baselineskip}%
  {0.5\baselineskip}%
  {\subsectitlefnt}}%
\newcommand{\ps@footnumber}{%
  \renewcommand{\@oddhead}{}%
  \renewcommand{\@evenhead}{}%
  \renewcommand{\@oddfoot}{\hfill{\thepage}}%
  \renewcommand{\@evenfoot}{{\thepage}\hfill}}
\makeatother

\newtheoremstyle{my_plain}%
{\parskip}
{\parskip}
{\normalfont\itshape}
{}
{\normalfont\thmfamily\bfseries}
{.}
{ }
{}

\newtheoremstyle{my_remark}%
{\parskip}
{\parskip}
{\normalfont}
{}
{\normalfont\thmfamily\bfseries}
{.}
{ }
{}

{
  \theoremstyle{my_plain}
  \newtheorem{theorem}{Theorem}[section]
  \newtheorem{proposition}[theorem]{Proposition}
  \newtheorem{corollary}[theorem]{Corollary}
  \newtheorem{lemma}[theorem]{Lemma}
  \newtheorem{definition}[theorem]{Definition}
}

{
  \theoremstyle{my_remark}
  \newtheorem{remark}[theorem]{Remark}
}

\numberwithin{equation}{section}

\newcommand{\fillbox}{\hspace*{\fill}\ensuremath{\Box}}

\newenvironment{abst}{\begin{quote}\small\thmfamily\bfseries
    Abstract. \normalfont\small}{\end{quote}}

%% file: math_commands.tex
\usepackage{amsmath}
\usepackage{amssymb}

\newcounter{mynumcounter}
\newenvironment{mynum}%
	       {%
		 \begin{list}{(\roman{mynumcounter})\hspace*{\fill}}%
		   {
		     \setlength{\topsep}{0cm}
		     \setlength{\partopsep}{0cm}
		     \setlength{\itemsep}{0ex}
		     \setlength{\parsep}{0cm}
		     \setlength{\leftmargin}{0cm}
		     \setlength{\itemindent}{8mm}		   
		     \setlength{\labelsep}{3mm}
		     \setlength{\labelwidth}{5mm}
		     \usecounter{mynumcounter}
		   }%
	       }
	       {\end{list}}

\newcommand{\eps}{\varepsilon}
\newcommand{\del}{\partial}

\newcommand{\dd}[2]{\frac{\del #1}{\del #2}}
\newcommand{\ddeval}[3]{\left.\dd{#1}{#2}\right|_{#3}}
\newcommand{\tdd}[2]{\tfrac{\del #1}{\del #2}}
\newcommand{\tddeval}[3]{\left.\tdd{#1}{#2}\right|_{#3}}

\newcommand{\Scal}{\operatorname{Scal}}

\newcommand{\tr}{\operatorname{tr}}
\renewcommand{\div}{\operatorname{div}}

\newcommand{\IR}{\mathbf{R}}

\newcommand{\CH}{\mathcal{H}}
\newcommand{\CL}{\mathcal{L}}
\newcommand{\CN}{\mathcal{N}}

\newcommand{\CX}{\mathcal{X}}

\newcommand{\id}{\operatorname{id}}

\newcommand{\rmd}{\mathrm{d}}

\newcommand{\supp}{\mathrm{supp}}

\newcommand{\tensor}{\otimes}
\newcommand{\dmu}{\,\rmd\mu}

\newcommand{\ra}{\rangle}
\newcommand{\la}{\langle}
\newcommand{\inj}{\mathrm{inj}}

\newcommand{\smallmathindent}{\setlength{\mathindent}{1em}\setlength{\multlinegap}{1em}}

\newcommand{\lap}{\Delta}
\newcommand{\Laplacian}[1]{\smash{\sideset{^{#1}}{}{\mathop\lap\nolimits}}}

\newcommand{\lapSig}{\Laplacian{\Sigma}}

\newcommand{\Connection}[1]{\smash{\sideset{^{#1}}{}{\mathop\nabla\nolimits}}}

\newcommand{\nabSig}{\!\Connection{\Sigma}}
\newcommand{\nabM}{\!\Connection{M}}
\newcommand{\nabL}{\!\Connection{L}}
\newcommand{\nabN}{\!\Connection{N}}

\newcommand{\Divergence}[1]{\smash{\sideset{^{#1}}{}{\mathop\mathrm{div}\nolimits}}}

\newcommand{\divM}{\!\Divergence{M}}

\newcommand{\Riemann}[1]{\smash{\sideset{^{#1}}{}{\mathop\mathrm{Rm}\nolimits}}}

\newcommand{\RiemL}{\!\Riemann{L}}
\newcommand{\RiemM}{\!\Riemann{M}}
\newcommand{\RiemSig}{\!\Riemann{\Sigma}}

\newcommand{\Ricci}[1]{\smash{\sideset{^{#1}}{}{\mathop\mathrm{Rc}\nolimits}}}

\newcommand{\RicL}{\!\Ricci{L}}
\newcommand{\RicM}{\!\Ricci{M}}
\newcommand{\RicSig}{\!\Ricci{\Sigma}}

\newcommand{\Scalarcurv}[1]{\smash{\sideset{^{#1}}{}{\mathop\mathrm{Sc}\nolimits}}}

\newcommand{\ScalL}{\!\Scalarcurv{L}}
\newcommand{\ScalM}{\!\Scalarcurv{M}}
\newcommand{\ScalSig}{\!\Scalarcurv{\Sigma}}

\newcommand{\Acircbar}%
{\hspace*{4pt}\raisebox{8.5pt}{\makebox[-4pt][l]{$\scriptstyle\circ$}}\bar A}
\newcommand{\Kcircbar}%
{\hspace*{4pt}\raisebox{8.5pt}{\makebox[-4pt][l]{$\scriptstyle\circ$}}\bar K}


%% file: my_commands.tex
\newcommand{\RicLpp}{\RicL(l^+\!\!,l^+\!)}
\newcommand{\RicLpm}{\RicL(l^+\!\!,l^-\!)}

\newcommand{\RiemLpmmp}{\RiemL(l^+\!\!,l^-\!\!,l^-\!\!,l^+\!)}

\newcommand{\IAk}{\int_{A(k)}\!\!}
\newcommand{\IAh}{\int_{A(h)}\!\!}

\newcommand{\two}{\mathrm{I\!I}}

\newcommand{\ds}{\,\rmd s}
\newcommand{\dl}{\,\rmd l}
\newcommand{\length}{\mathrm{length}}

%% file: intro.tex
 \newcommand{\mnote}[1]{\marginpar{\raggedright\tiny\em
 $\!\!\!\!\!\!\,\bullet$ #1}}       
\section{Introduction}
\label{sec:introduction}
The celebrated regularity result for stable minimal surfaces, due to
Schoen, Simon, and Yau \cite{Schoen-Simon-Yau:1975}, gives a bound on
the second fundamental form in terms of ambient curvature and area of
the surface.  The proof of the main result of
\cite{Schoen-Simon-Yau:1975} makes use of the Simons formula
\cite{Simons:1968} for the Laplacian of the second fundamental form,
together with the non-negativity of the second variation of area. 
In this paper we will prove a
generalization of the regularity result of Schoen, Simon, and Yau to
the natural analogue of stable minimal surfaces in the context of
Lorentz geometry, stable marginally trapped surfaces. In this case, a
generalization of the Simons formula holds for the null second fundamental
form, and the appropriate notion of stability is that of stably outermost in
the sense of \cite{Andersson-Mars-Simon:2005,Newman:1987}. A  
local area estimate for stable marginally trapped surfaces, a generalization
of a result due to Pogorelov \cite{Pogorelov:1981} allows us to give
a curvature bound independent of assumptions on the area of the surface.

Let $\Sigma$ be a spacelike surface of co-dimension two in a 3+1
dimensional Lorentz manifold $L$ and let $l^{\pm}$ be the two
independent future directed null sections of the normal bundle of
$\Sigma$, with corresponding mean curvatures, or null expansions,
$\theta^{\pm}$. $\Sigma$ is called trapped if the future directed
null rays starting at $\Sigma$ converge, {\em i.e.} $\theta^{\pm} <
0$. If $L$ contains a trapped surface and satisfies certain causal
conditions, then, if in addition, the null energy condition is
satisfied, $L$ is future causally incomplete \cite{Penrose:1965}. Let
$l^+$ be the outgoing null normal. If $L$ is an asymptotically flat
spacetime this notion is well defined, otherwise the outgoing
direction can be fixed by convention.  We call $\Sigma$ a marginally
outer trapped surface (MOTS) if the outgoing lightrays are marginally
converging, {\em i.e.} if $\theta^+ = 0$. No assumption is made on the
ingoing null expansion $\theta^-$ of a MOTS. If $\Sigma$ is contained in
a time symmetric Cauchy surface, then $\theta^+ = 0$ if and only if
$\Sigma$ is minimal.

Marginally trapped surfaces are of central importance in general
relativity, where they play the role of apparent horizons, or
quasilocal black hole boundaries. The conjectured Penrose inequality,
proved in the Riemannian case by Huisken and Ilmanen
\cite{Huisken-Ilmanen:2001} and Bray \cite{Bray:2001}, may be
formulated as an inequality relating the area of the outermost
apparent horizon and the ADM mass.  The technique of excising the
interior of black holes using apparent horizons as excision boundaries
plays a crucial role in current work in numerical relativity, where
much of the focus is on modelling binary black hole collisions.

In spite of the importance of marginally trapped surfaces in the
geometry of spacetimes, the extent of our knowledge of the regularity
and existence of these objects is rather limited compared to the
situation for minimal surfaces.

A smooth marginally outer trapped
surface is stationary with respect to variations of 
area within its outgoing null cone, in view of the formula
\begin{equation*}
  \delta_{f l^+} \mu_\Sigma = f  \theta^+ \mu_\Sigma
\end{equation*}
where $f$ is a function on $\Sigma$. The second variation of area at a
MOTS in the direction $l^+$ is
\begin{equation*}
  \delta_{f l^+} \theta^+ = - (|\chi^+|^2 + G(l^+, l^+)) f
\end{equation*}
where $G$ denotes the Einstein tensor of $L$, and $\chi^+$ is the
second fundamental form of $\Sigma$ with respect to $l^+$.  For
minimal surfaces in a Riemannian manifold, or maximal hypersurfaces in
a Lorentz manifold the second variation operator is an elliptic
operator of second order. In contrast, the above equation shows that
the second variation operator for area of a MOTS, with respect to
variations in the null direction $l^+$, is an operator of order zero.
Therefore, although MOTS can be characterized as stationary points of
area, this point of view alone is not sufficient to yield a useful
regularity result.  In spite of this, as we shall see below, there is a
natural generalization of the stability condition for minimal
surfaces, as well as of the regularity result of Schoen, Simon, and Yau,
to marginally outer trapped surfaces.

It is worth remarking at this point that if we consider variations of
area of spacelike hypersurfaces in a Lorentz manifold, the stationary
points are maximal surfaces. Maximal surfaces satisfy a quasilinear
non-uniformly elliptic equation closely related to the minimal surface
equation. Due to the fact that maximal hypersurfaces are
spacelike, they are Lipschitz submanifolds. Moreover, in a spacetime
satisfying the timelike convergence condition, every maximal surface
is stable. Hence, the regularity theory for maximal surfaces is of a
different flavor than the regularity theory for minimal
surfaces, cf.~\cite{Bartnik:1984}.

Assume that $L$ is provided with a reference foliation consisting of
spacelike hyper\-surfaces~$\{M_t\}$, and that $\Sigma$ is contained in
one of the leaves $M$ of this foliation. Let $(g,K)$ be the induced
metric and the second fundamental form of $M$ with respect to the
future directed timelike normal $n$. Further, let $\nu$ be the
outward pointing normal of $\Sigma$ in $M$ and let $A$ be the second
fundamental form of $\Sigma$ with respect to $\nu$. After possibly
changing normalization, $l^{\pm} = n \pm \nu$, we have
\begin{equation*}
  \theta^{\pm} = H \pm \tr_\Sigma K
\end{equation*}
where $H = \tr A$ is the mean curvature of $\Sigma$ and $\tr_\Sigma K$ is the
trace of the projection of $K$ to $\Sigma$. Thus the condition for $\Sigma$
to be a MOTS, $\theta^+ = 0$, is a prescribed mean curvature
equation.

The condition that plays the role of stability for MOTS is the stably
outermost condition,
see~\cite{Andersson-Mars-Simon:2005,Newman:1987}. Suppose $\Sigma$ is
contained in a spatial hypersurface $M$. Then $\Sigma$ is stably
locally outermost in $M$ if there is an outward infinitesimal deformation of
$\Sigma$, within $M$ which does not decrease $\theta^+$. This
condition, which is equivalent to the condition that $\Sigma$ is
stable in case $M$ is time symmetric, turns out to be sufficient to
apply the technique of~\cite{Schoen-Simon-Yau:1975} to prove a bound
on the second fundamental form $A$ of $\Sigma$ in $M$. In contrast to the
situation for minimal surfaces the stability operator defined by the
deformation of $\theta^+$ is not self-adjoint. Nevertheless, it has a real
principal eigenvalue with a corresponding principal eigenfunction which does
not change sign. 

The techniques of \cite{Schoen-Simon-Yau:1975} were first applied in
the context of general relativity by Schoen and Yau
\cite{Schoen-Yau:1981}, where existence and regularity for Jang's
equation were proved. Jang's equation is an equation for a graph in $N
= M \times \IR$, and is of a form closely related to the equation
$\theta^+ = 0$. Let $u$ be a function on $M$, and let $\bar K$ be the
pull-back to $N$ of $K$ along the projection $N \to M$.  Jang's
equation is the equation
\begin{equation*}
  \bar g^{ij} \left ( \frac{D_i D_j u}{\sqrt{1 + |Du|^2}} + \bar
  K_{ij} \right ) = 0
\end{equation*}
where $\bar g^{ij} = g^{ij} - \frac{D_i u D_j u}{1+|Du|^2}$ is the induced
metric on the graph $\bar \Sigma$ of $u$ in $N$.  
Thus Jang's equation can be written as $\bar \theta= 0$ with
\begin{equation*}
\bar \theta = \bar H + \tr_{\bar \Sigma} \bar K ,
\end{equation*}
where $\bar H$ is the mean curvature of $\bar \Sigma$ in $N$. This shows
that Jang's equation $\bar \theta = 0$ 
is a close analog to the equation $\theta^+ = 0$ characterizing a
MOTS. 
Solutions to Jang's equation satisfy a stability condition closely
related to the stably outermost condition stated above. This
is due to the fact that Jang's equation is translation invariant in
the sense that if $u$ solves Jang's equation, then also $u + c$ is a
solution where $c$ is a constant. Thus, in the sense of section
\ref{sec:stable}, graphical solutions to Jang's equation are stable.
This fact allows Schoen and Yau~\cite{Schoen-Yau:1981} to apply the
technique of~\cite{Schoen-Simon-Yau:1975} to prove regularity for
solutions of Jang's equation.  
It is worth remarking that
although the dominant energy condition is assumed to hold
throughout~\cite{Schoen-Yau:1981}, in fact the proof of the existence
and regularity result for solutions of Jang's equation presented
in~\cite{Schoen-Yau:1981} can be carried out without this assumption.
In the present paper, the dominant energy condition is not used in the
proof of our main regularity result, cf. Theorem~\ref{thm:main-intro}
below.

It was proved by Galloway and Schoen \cite{Galloway-Schoen:2005},
based on an argument for solutions of the Jang's equation
in~\cite{Schoen-Yau:1981}, that the stability of MOTS implies a
``symmetrized'' stability condition, which states that the spectrum of
a certain self-adjoint operator analogous to the second variation
operator for minimal surfaces is non-negative. The fact that stability
in the sense of stably outermost implies this symmetrized version of
stability was used in~\cite{Galloway-Schoen:2005} to give conditions
on the Yamabe type of stable marginal surfaces in general dimension.
It turns out that this weaker symmetrized notion of stability is in
fact sufficient for the curvature estimates proved here. The
symmetrized notion of stability is also used in our local area
estimates. However, since this notion has no direct interpretation in
terms of the geometry of the ambient spacetime, we prefer to state our
results in terms of the stably outermost condition.



\subsection{Statement of Results}
The stability condition for MOTS which replaces the stability condition for
minimal surfaces and which allows one to apply the technique of
\cite{Schoen-Simon-Yau:1975} is the following. 
\begin{definition} 
  \label{def:stable-intro} 
  $\Sigma$ is stably outermost if there
  is a function $f \geq 0$ on $\Sigma$, $f \ne 0$ somewhere, such that
  $\delta_{f \nu} \theta^+ \geq 0$.
\end{definition} 

When there is no room for confusion we will refer to a stably outermost MOTS
simply as a stable MOTS. 
This is analogous to the stability condition for a minimal surface $N \subset
M$. The condition that there exist a function $f$ on $N$, $f \geq 0$, $f \ne
0$ somewhere, such that $\delta_{f \nu} H \geq 0$ is equivalent to the
condition that $N$ is stable. 

The main result of this paper is the following theorem, cf. theorem
\ref{thm:main}, corollary \ref{cor:main} as well as theorem
\ref{thm:area_bound}.
\begin{theorem}
  \label{thm:main-intro} 
  Suppose $\Sigma$ is a stable MOTS in $(M,g,K)$.  Then the second
  fundamental form $A$ satisfies the inequality
  \begin{equation*}
  \|A\|_\infty \leq C(\|K\|_\infty,\|\nabla K\|_\infty,
  \|\RiemM\|_\infty, \mathrm{inj}(M,g)^{-1} )\,.
  \end{equation*}
\end{theorem} 
Here $\|\cdot\|_{\infty}$ denotes the $\sup$-norm of the respective
quantity, taken on $\Sigma$. As an application we prove a compactness
result for MOTS, cf. theorem \ref{thm:compact}.
\begin{theorem}
  \label{thm:compact-intro}
  Let $(g_n, K_n)$ be a sequence of initial data sets on a manifold
  $M$. Let $(g,K)$ be another initial dataset on $M$
  such that
  \begin{align*}
    \|\RiemM\|_{\infty}&\leq C\,,
    \\[.2ex]
    \|K\|_{\infty} + \|\nabM K\|_{\infty} &\leq C, 
    \\[.2ex]
    \mathrm{inj} (M,g) &\geq C^{-1}\,,
    \intertext{for some constant $C$. Assume that}
    g_n \to g \quad&\text{in}\quad C_\text{loc}^2(M,g)\ \text{and},\\
    K_n \to K \quad&\text{in}\quad C_\text{loc}^1(M,g).
  \end{align*}
  Furthermore, let $\Sigma_n\subset M$ be a sequence of immersed
  surfaces which are stable marginally outer trapped with respect to
  $(g_n, K_n)$ and have an accumulation point in $M$. In addition,
  assume that the $\Sigma_n$ have uniformly locally finite area, that
  is, for all $x\in M$ there exists $0<r=r(x)$ and $a=a(x)<\infty$
  such that
  \begin{equation*}
    |\Sigma_n \cap B_{M_{t_n}}(x,r)| \leq ar^2 \qquad
    \text{uniformly in\ } n,
  \end{equation*}
  where $B_{M_{t_n}}(x,r)$ denotes the ball in $M$ around $x$ with
  radius $r$.

  Then a subsequence of the $\Sigma_n$ converges to a smooth immersed
  surface $\Sigma$ locally in the sense of $C^{1,\alpha}$ graphs.
  $\Sigma$ is a MOTS with respect to $(g, K)$. If $\Sigma$ is
  compact, then it is also stable.  \fillbox
\end{theorem}
\subsection{Outline of the paper}
In sections \ref{sec:prelim} and \ref{sec:simons} we discuss the
notation and preliminary results, as well as a Simons identity which
holds for the shear of a MOTS. Section \ref{sec:linearization}
introduces the linearization of the operator $\theta^+$ acting on
surfaces represented as graph over a MOTS. The stability conditions we
use are discussed in section \ref{sec:stable}. The curvature 
estimates are derived in section \ref{sec:apriori} under the
assumption of local area bounds. In section \ref{sec:area} we show how
these bounds can be derived in terms of the ambient geometry. Finally
section \ref{sec:applic} uses the established curvature bounds to
prove the compactness theorem.


%% file: prelim.tex
\section{Preliminaries and Notation}
\label{sec:prelim}
In this section we set up notation and recall some preliminaries from
differential geometry.  In the sequel we will consider two-dimensional
spacelike submanifolds $\Sigma$ of a four dimensional manifold $L$. As
a space time manifold, $L$ will be equipped with a metric $h$ of
signature $(-,+,+,+)$.  The inner product induced by $h$ will
frequently be denoted by $\la\cdot,\cdot\ra$. In addition, we will
assume, that $\Sigma$ is contained in a spacelike hypersurface $M$ in
$L$. The metric on $M$ induced by $h$ will be denoted by $g$, the
metric on $\Sigma$ by $\gamma$. We will denote the tangent bundles by
$TL,TM$, and $T\Sigma$, and the space of smooth tangential vector
fields along the respective manifolds by $\CX(\Sigma)$, $\CX(M)$, and
$\CX(L)$. Unless otherwise stated, we will assume that all manifolds
and fields are smooth. 

We denote by $n$ the future directed unit timelike normal of $M$ in
$L$, which we will assume to be a well defined vector field along
$M$. The normal of $\Sigma$ in $M$ will be denoted by $\nu$, which
again is assumed to be a well defined vector field along $\Sigma$.

The two directions $n$ and $\nu$ span the normal bundle $\CN\Sigma$ of $\Sigma$ in
$L$, and moreover, we can use them to define two canonical null
directions, which also span this bundle, namely $l^\pm := n\pm \nu$.

In addition to the metrics, $h$ and its Levi-Civita connection $\nabL$
induce the second fundamental form $K$ of $M$ in $L$. It is the
normal part of $\nabL$, in the sense that for all vector fields
$X,Y\in\CX(M)$
\begin{equation}
  \label{eq:secondff_ML}
  \nabL_X Y = \nabM_X Y + K(X,Y) n\,.
\end{equation}
The second fundamental form of $\Sigma$ in $M$ will be denoted by $A$.
For vector fields $X,Y\in\CX(\Sigma)$ we have
\begin{equation}
  \label{eq:secondff_SM}
  \nabM_X Y = \nabSig_X Y - A(X,Y) \nu\,.
\end{equation}
For vector fields $X,Y\in\CX(\Sigma)$, the
connection of $L$ therefore splits according to
\begin{equation}
  \label{eq:secondff_SL}
  \nabL_X Y = \nabSig_X Y + K^\Sigma(X,Y) n - A(X,Y) \nu = \nabSig_X Y -
  \two(X,Y)\,,
\end{equation}
where $\two(X,Y) = A(X,Y)\nu - K^\Sigma(X,Y) n$ is the second fundamental
form of $\Sigma$ in $L$. Here $K^\Sigma$ denotes the restriction of
$K$ to $T\Sigma$, the tangential space of $\Sigma$.

The trace of $\two$ with respect to $\gamma$, which is a vector in the
normal bundle of $\Sigma$, is called the mean curvature vector and is
denoted by
\begin{equation}
  \label{eq:mean_curv_vect}
  \CH = \sum_{i} \two(e_i,e_i)\,,
\end{equation}
for an orthonormal basis $e_1,e_2$ of $\Sigma$. Since $\CH$ is normal
to $\Sigma$, it satisfies
\begin{equation}
  \label{eq:H_split}
  \CH= H\nu - P n
\end{equation}
where $H=\gamma^{ij}A_{ij}$ is the trace of $A$ and
$P=\gamma^{ij}K^\Sigma_{ij}$ the trace of
$K^\Sigma$, with respect to $\gamma$.
For completeness, we note that the norms of $\two$ and $\CH$ are given
by
\begin{align}
  \label{eq:norm_two}
  |\two|^2 &= |A|^2-|K^\Sigma|^2\qquad\text{and}
  \\
  \label{eq:norm_mcv}
  |\CH|^2 &= H^2-P^2\,.
\end{align}
Recall that since $\CH$ and $\two$ have values normal to $\Sigma$, the
norms are taken with respect to $h$ and are therefore not necessarily nonnegative.

We use the following convention to represent the Riemann curvature
tensor $\RiemSig$, the Ricci tensor $\RicSig$, and the scalar curvature
$\ScalSig$ of $\Sigma$. Here $X,Y,U,V\in\CX(\Sigma)$ are vector fields.
\begin{align*}
  &\RiemSig(X,Y,U,V) = \big\la \nabSig_X\nabSig_Y U - \nabSig_Y\nabSig_X U -
  \nabSig_{[X,Y]} U,V \big\ra\,,\phantom{\sum_i}
  \\
  &\RicSig(X,Y) =  \sum_i \RiemSig(X,e_i,e_i,Y)\,,  
  \\
  &\ScalSig = \sum_i \RicSig(e_i,e_i)\,.
\end{align*}
Analogous definitions hold for $\RiemM,\RicM,\ScalM$ and
$\RiemL,\RicL,\ScalL$, with the exception that for $\RicL$ and
$\ScalL$ we take the trace with respect to the indefinite metric $h$.

We recall the Gauss and Codazzi equations of $\Sigma$ in $L$, which
relate the respective curvatures. The Riemann curvature tensors
$\RiemSig$ and $\RiemL$ of $\Sigma$ and $L$ respectively, are related
by the Gauss equation. For vector fields $X,Y,U,V$ we have
\begin{multline}
  \label{eq:gauss}
  \RiemSig(X,Y,U,V) =\\
  \RiemL(X,Y,U,V) + \big\la\two(X,V),\two(Y,U)\big\ra
  - \big\la\two(X,U),\two(Y,V)\big\ra\,.
\end{multline}
In two dimensions, all curvature information of $\Sigma$ is contained
in its scalar curvature, which we will denote by $\ScalSig$. The scalar
curvature of $L$ will be denoted by $\ScalL$. The information of the
Gauss equation above is fully contained in the following equation,
which emerges from the above one by first taking the trace with
respect to $Y,U$ and then with respect to $X,V$
\begin{equation}
  \label{eq:gauss_tr}
  \ScalSig = \ScalL + 2\RicL(n,n) - 2 \RicL(\nu,\nu)
  -2\RiemL(\nu,n,n,\nu) + |\CH|^2 - |\two|^2\,.
\end{equation}
The Codazzi equation, which relates $\RiemL$ to $\two$, has the
following form
\begin{equation}
  \label{eq:codazzi}
  \big\la\nabL_X \two(Y,Z), S\big\ra = \big\la\nabla_Y \two(X,Z),S\big\ra
  + \RiemL(X,Y,S,Z)
\end{equation}
for vector fields $X,Y,Z\in\CX(\Sigma)$ and $S\in\Gamma(\CN \Sigma)$.

There is also a version of the Gauss and Codazzi equations for the
embedding of $M$ in $L$. They relate the curvature $\RiemL$ of $L$ to
the curvature $\RiemM$ of $M$. For vector fields $X,Y,U,V\in\CX(M)$ we
have
\begin{gather}
  \label{eq:gauss_ML}
  \begin{split}
    &\RiemM(X,Y,U,V) 
    \\
    &\quad = \RiemL(X,Y,U,V) - K(Y,U)K(X,V) + K(X,U)K(Y,V)\,,
  \end{split}
  \\
  \label{eq:codazzi_ML}
  \nabM_X K(Y,U) - \nabM_Y K(X,U) =  \RiemL(X,Y,n,U)\,.
\end{gather}
These equations also have a traced form, namely
\begin{gather}
  \label{eq:gauss_ML_tr}
  \ScalM = \ScalL + 2\RicL(n,n) - (\tr K)^2 + |K|^2\,\ \text{and} \\
  \label{eq:codazzi_ML_tr}
  \divM K - \nabM \tr K = \RicL(\cdot,n)\,.
\end{gather}
We now investigate the connection $\nabN$ on the normal bundle $\CN\Sigma$ of
$\Sigma$. Recall that for sections $N$ of $\CN\Sigma$ and
$X\in\CX(\Sigma)$, this connection is defined as follows 
\begin{equation*}
  \nabN_X N = \big(\nabL_X N\big)^\perp\,,
\end{equation*}
where again $(\cdot)^\perp$ means taking the normal part.
We have
\begin{equation*}
  0 = X(1) = X \big( \la n,n \ra \big)  = 2 \la \nabN_X n,n \ra\,,
\end{equation*}
and similarly $\la\nabN_X \nu,\nu\ra =0$. Therefore the relevant component
of $\nabN$ is
\begin{equation*}
  \big\la\nabN_X \nu,n\big\ra = \big\la\nabL_X \nu,n\big\ra = - K(X,\nu)\,.
\end{equation*}
Recall that $X$ is tangential to $\Sigma$. This lead us to define the 
1-form $S$ along $\Sigma$ by the restriction of $K(\cdot,\nu)$ to
$T\Sigma$.
\begin{equation}
  \label{eq:def_S}
  S(X) := K(X,\nu)\,.
\end{equation}
Then, for an arbitrary section $N$ of $\CN\Sigma$ with $N=f\nu+g n$,  we have 
\begin{equation*}
  \nabN_X N = X(f)\nu + X(g)n + S(X)\big(fn + g\nu)\,.
\end{equation*}
In particular 
\begin{equation}
  \label{eq:nab_l}
  \nabN_X l^\pm = \pm S(X) l^\pm\,.
\end{equation}
We will later consider the decomposition of $\two$ into its null
components. For $X,Y\in \CX(\Sigma)$ let
\begin{equation}
  \label{eq:null_ff}
  \chi^\pm(X,Y):= \big\la\two (X,Y),l^\pm\ra =  K(X,Y) \pm A(X,Y)\,.
\end{equation}
The traces of $\chi^\pm$ respectively will be called $\theta^\pm$
\begin{equation}
  \label{eq:theta_pm}
  \theta^\pm = \la\CH,l^\pm\ra = P \pm H\,.
\end{equation}
The Codazzi-equation \eqref{eq:codazzi} implies a Codazzi equation for $\chi^\pm$.
\begin{lemma}
  For vector fields $X,Y,Z\in\CX(\Sigma)$ the following relation holds
  \begin{equation}
    \label{eq:codazzi_null}
    \nabla_X \chi^\pm(Y,Z) = \nabla_Y \chi^\pm(X,Z) + Q^\pm(X,Y,Z) \mp
    \chi^\pm(X,Z)S(Y) \pm \chi^\pm(Y,Z)S(X)\,.
  \end{equation}
  Here
  \begin{equation}
    \label{eq:Qdef}
    Q^\pm(X,Y,Z) = \RiemL(X,Y,l^\pm,Z).
  \end{equation}
\end{lemma}

%% file: simons.tex
\section{A Simons identity for $\chi^\pm$}
\label{sec:simons}
We use the Codazzi equation we derived in the previous section to
compute an identity for the Laplacian of $\chi^\pm$, which is very
similar to the Simons identity for the second fundamental form of a
hypersurface \cite{Simons:1968, Schoen-Simon-Yau:1975}.

The Laplacian on the surface $\Sigma$ is defined as the operator
\[ \lapSig = \gamma^{ij} \nabSig^2_{ij}\,. \]
In the sequel, we will drop the superscript on $\lapSig$ and $\nabSig$,
since all tensors below will be defined only along $\Sigma$. We will
switch to index notation, since this is convenient for the
computations to follow. In this notation  
\begin{equation*}
  T^{i_1\cdots i_p}_{j_1\cdots j_q}
\end{equation*}
denotes a $(p,q)$-tensor $T$ as the collection of its components in
an arbitrary basis $\{\del_i\}_{i=1}^2$ for the tangent spaces. To
make the subsequent computations easier, we will usually pick a basis of
normal coordinate vectors. Also note that we use Latin indices ranging
from 1 to 2 to denote components tangential to the surface $\Sigma$. 

Recall, that the commutator of the connection is given by the Riemann
curvature tensor, such that for a $(0,2)$-tensor $T_{ij}$
\begin{equation}
  \label{eq:riem_2tens} 
  \nabla_k\nabla_l T_{ij} - \nabla_l \nabla_k T_{ij} = \RiemSig_{klmi}
  T_{mj} + \RiemSig_{klmj} T_{im}\,.
\end{equation}
Note that we use the shorthand $\RiemSig_{klmj} T_{im} = \RiemSig_{klpj}
T_{iq}\gamma^{pq}$, when there is no ambiguity. That is, we assume
that we are in normal coordinates where
$\gamma_{ij}=\gamma^{ij}=\delta_{ij}$.
Also note that this fixes the sign
convention for $\RiemSig_{ijkl}$ such that $\RicSig_{ij}= \RiemSig_{ikkj}$
is positive on the round sphere.
\begin{lemma}
  \label{lemma:simons}
  The Laplacian of $\chi=\chi^+$ satisfies the following identity
  \begin{equation*}
    \begin{split}
    \chi_{ij}\lap\chi_{ij} &= \chi_{ij} \nabla_i\nabla_j
      \theta^+ + \chi_{ij}\big(\RiemL_{kilk}\chi_{lj} +
      \RiemL_{kilj}\chi_{kl}\big)
      \\ & \phantom{=}
      + \chi_{ij} \nabla_k \big(Q_{kij} - \chi_{kj}S_i +
      \chi_{ij} S_k\big)
      + \chi_{ij} \nabla_i \big(Q_{kjk} - \theta^+ S_j + \chi_{jk} S_k\big)
      \\ & \phantom{=}
      - |\two|^2 |\chi|^2
      + \theta^+ \chi^+_{ij}\chi^+_{jk}\chi^+_{ki}
      - \theta^+ \chi^+_{ij}\chi^+_{jk} K^\Sigma_{ki}
      - P\chi^+_{ij}\chi^+_{jk}\chi^+_{ki}
    \end{split}
  \end{equation*}
  where $P=\gamma^{ij}K^\Sigma_{ij}$ is the trace of $K^\Sigma$.
\end{lemma}
\begin{proof}
  Recall that in coordinates the Codazzi equation
  \eqref{eq:codazzi_null} for $\chi_{ij}$
  reads
  \begin{equation}
    \label{eq:codazzi_coord}
    \nabla_i\chi_{jk} = \nabla_j\chi_{ik} + Q_{ijk} -
    \chi_{ik}S_j +  \chi_{jk}S_i\,.
  \end{equation}
  Then compute, using \eqref{eq:codazzi_coord} in the first and third
  step, and the commutator relation \eqref{eq:riem_2tens} in the
  second, to obtain
  \begin{equation}
    \label{eq:simons_1}
    \begin{split}
      \nabla_k\nabla_l \chi_{ij}
      &= \nabla_k \nabla_i\chi_{lj}
      + \nabla_k \big(Q_{lij} - \chi_{lj}S_i + \chi_{ij}S_l\big)
      \\
      &= \nabla_i\nabla_k\chi_{lj} + \RiemSig_{kiml}\chi_{mj} +
      \RiemSig_{kimj}\chi_{lm} 
      \\ & \phantom{=}
      + \nabla_k \big(Q_{lij} - \chi_{lj}S_i + \chi_{ij}S_l\big)
      \\
      &= \nabla_i\nabla_j \chi_{kl} + \RiemSig_{kiml}\chi_{mj} +
      \RiemSig_{kimj}\chi_{lm} 
      \\ & \phantom{=}
      + \nabla_k \big(Q_{lij} - \chi_{lj}S_i + \chi_{ij} S_l\big)
      + \nabla_i \big(Q_{kjl} - \chi_{kl}S_j + \chi_{jl} S_k\big)\,.
    \end{split}    
  \end{equation}
  We will use the Gauss equation \eqref{eq:gauss} to replace the
  $\RiemSig$-terms by $\RiemL$-terms. Observe, that
  \begin{equation*}
    \two_{ij} = -\tfrac{1}{2} \chi^+_{ij} l^- - \tfrac{1}{2}
    \chi^-_{ij} l^+\,.
  \end{equation*}
  Plugging this into the Gauss equation \eqref{eq:gauss} gives
  \begin{equation*}
    \RiemSig_{ijkl} = \RiemL_{ijkl} + \tfrac{1}{2}\big(
    \chi^+_{ik}\chi^-_{jl} + \chi^-_{ik}\chi^+_{jl} -
    \chi^+_{il}\chi^-_{jk} - \chi^-_{il}\chi^+_{jk} \big)\,.
  \end{equation*}
  Combining with \eqref{eq:simons_1}, we infer that
  \begin{equation*}
    \begin{split}
      \nabla_k\nabla_l \chi_{ij}
      &= \nabla_i\nabla_j \chi_{kl} + \RiemL_{kiml}\chi_{mj} +
      \RiemL_{kimj}\chi_{lm}
      \\ &\phantom{=}
      + \tfrac{1}{2}\big( 
      \chi^+_{il}\chi^-_{km} + \chi^-_{il}\chi^+_{km} 
      - \chi^+_{kl}\chi^-_{im} - \chi^-_{kl}\chi^+_{im}
      \big)\chi^+_{mj}
      \\ &\phantom{=}
      + \tfrac{1}{2}\big(
      \chi^+_{km}\chi^-_{ij} + \chi^-_{km}\chi^+_{ij}
      - \chi^+_{kj}\chi^-_{im} - \chi^-_{kj}\chi^+_{im}
      \big)\chi^+_{lm}
      \\ & \phantom{=}
      + \nabla_k \big(Q_{lij} - \chi_{lj}S_i + \chi_{ij} S_l\big)
      + \nabla_i \big(Q_{kjl} - \chi_{kl}S_j + \chi_{jl} S_k\big)\,.
    \end{split}
  \end{equation*}
  Taking the trace with respect to $k,l$ yields
  \begin{equation*}
    \begin{split}
      \lap \chi_{ij} &= \nabla_i\nabla_j \theta^+ +
      \RiemL_{kilk}\chi_{lj} + \RiemL_{kilj}\chi_{kl} 
      \\ & \phantom{=}
      + \nabla_k \big(Q_{kij} - \chi_{kj}S_i + \chi_{ij} S_k\big)
      + \nabla_i \big(Q_{kjk} - \theta^+ S_j + \chi_{jk} S_k\big)
      \\ & \phantom{=}
      + \frac{1}{2} \big(
      \chi^-_{ij} |\chi^+|^2 + \la \chi^+,\chi^- \ra
      \chi^+_{ij}
      -\theta^+\chi^+_{jk}\chi^-_{ki} -
      \theta^-\chi^{+}_{jk}\chi^+_{ki}
      \big)
      \\ & \phantom{=}
      + \frac{1}{2} \big( 
      \chi^+_{jk}\chi^-_{kl}\chi^+_{li} -
      \chi^-_{jk}\chi^+_{kl}\chi^+_{li} \big)
    \end{split}
  \end{equation*}
  We contract this equation with $\chi^+_{ij}$ and obtain
  \begin{equation*}
    \begin{split}
      \chi_{ij}\lap\chi_{ij} &= \chi_{ij} \nabla_i\nabla_j
      \theta^+ + \chi_{ij}\big(\RiemL_{kilk}\chi_{lj} +
      \RiemL_{kilj}\chi_{kl}\big)
      \\ & \phantom{=}
      + \chi_{ij} \nabla_k \big(Q_{kij} - \chi_{kj}S_i +
      \chi_{ij} S_k\big)
      + \chi_{ij} \nabla_i \big(Q_{kjk} - \theta^+ S_j + \chi_{jk} S_k\big)
      \\ & \phantom{=}
      + \la \chi^+, \chi^-\ra |\chi|^2
      - \tfrac{1}{2} \theta^+ \chi^+_{ij}\chi^+_{jk}\chi^-_{ki} 
      - \tfrac{1}{2} \theta^- \chi^+_{ij}\chi^+_{jk}\chi^+_{ki}\,.
    \end{split}
  \end{equation*}
  Now observe that $\chi^-_{ij} = 2K^\Sigma_{ij} - \chi^+_{ij}$
  and $\theta^- = 2P -\theta^+$. Substituting this into the last two
  terms, together with $\la \chi^+, \chi^-\ra = -|\two|^2$, we
  arrive at the identity we claimed.
\end{proof}

%% file: linearization.tex
\section{The Deformation of $\theta^+$}
\label{sec:linearization}
This section is concerned with the deformation of the operator
$\theta^+$, as defined in equation \eqref{eq:theta_pm}. We begin by
considering an arbitrary, spacelike hypersurface $\Sigma\subset L$.
Assume that the normal bundle is spanned by the globally defined null
vector fields $l^\pm$, such that $\la l^+,l^-\ra=-2$. We call such a frame
a \emph{normalized null frame}. As before, let
$\theta^\pm:=\la\CH,l^\pm\ra$. We abbreviate $\chi=\chi^+$.

A variation of $\Sigma$ is a differentiable map 
\[ F:\Sigma \times (-\eps,\eps)\to L : (x,t) \mapsto F(x,t)\,, \]
such that $F(\cdot,0)=\id_\Sigma$ is the identity on $\Sigma$. The
vector field $\ddeval{F}{t}{t=0}=V$ is called variation vector field
of $F$. We will only consider variations, with variation vector fields
$V$ of the form $V = \alpha l^+  + \beta l^-$.

Note that in this setting, as a normalized null frame is not uniquely
defined by its properties, the notion of $\theta^+$ depends on the
frame chosen. The freedom we have here is the following. Assume
$k^\pm$ is another normalized null frame for the normal bundle of
$\Sigma$, that is $h(k^\pm,k^\pm)=0$ and $h(k^+,k^-)=-2$. Since the
null cone at each point is unique, the directions of $k^\pm$ can be
aligned with $l^\pm$. But their magnitudes can be different, so $k^+ = e^\omega l^+$
and $k^- = e^{-\omega} l^-$ with a function $\omega\in C^\infty(\Sigma)$.

Therefore, if we want to compute the deformation of $\theta^+$, it
will not only depend on the deformation of $\Sigma$, as encoded in the
deformation vector $V$. It will also depend on the change of the
frame, that is on the change of the vector $l^+$, which is an
additional degree of freedom.

To expose the nature of that freedom, observe that if $l^\pm(t)$ is a null frame
on each $\Sigma_t := F(\Sigma,t)$, then $\ddeval{l^\pm}{t}{t=0}$ is still
normal to $\Sigma$. On the other hand 
\begin{gather*}
  0 = \tddeval{}{t}{t=0} \la l^+,l^+\ra =
  2\big\la\tddeval{l^+}{t}{t=0},l^+\big\ra \qquad\text{and} \\
  0= \tddeval{}{t}{t=0} \la l^+,l^-\ra =
  \big\la\tddeval{l^+}{t}{t=0},l^-\big\ra +
  \big\la\tddeval{l^-}{t}{t=0},l^+\big\ra
\end{gather*}
Therefore $\ddeval{l^\pm}{t}{t=0} = w l^\pm$ for a function
$w\in C^\infty(\Sigma)$. Thus the linearized change of the frame
is described by the single function $w$, which we will call the
\emph{variation of the null frame}.

If we fix both of the quantities $V$ and $w$, a straight forward
(but lengthy) computation gives the deformation of $\theta^+$.
\begin{lemma}
  \label{lemma:linearization}
  Assume $F:\Sigma\times (-\eps,\eps)\to L$ is a variation of $\Sigma$
  with variation vector field $V = \alpha l^+ + \beta l^-$. Assume
  further that the variation of the null frame is $w$. Then the
  variation of $\theta^+$ is given by 
  \begin{equation*}
    \begin{split}
      \delta_{V,w}\theta^+ 
      &= 2 \Delta \beta - 4 S(\nabla\beta)
      - \alpha \big( |\chi|^2 + \RicLpp \big) + 2 \theta^+ w
      \\ 
      &\phantom{=}
      - \beta \big( 2 \div S - 2 |S|^2 - |\two|^2 + \RicLpm 
      - \tfrac{1}{2} \RiemLpmmp \big) \,.
    \end{split}
  \end{equation*}
\end{lemma}
If we consider marginally trapped surfaces, then the term $\theta^+w$
in the previous calculation vanishes, and we get expressions
independent of the change in the frame. As a consequence, we state
the following two corollaries, which also restrict the variations we
take into account.
\begin{corollary}
  \label{cor:Lminus}
  Assume $\Sigma$ is a marginally trapped surface, that is, it
  satisfies the equation $\theta^+=0$. Then the deformation of
  $\theta^+$ in direction of $-l^-$ is given by 
  \begin{equation*}
    \delta_{-\beta l^-,w} \theta^+ = 2L_- \beta\,,
  \end{equation*}
  where the operator $L_-$ is given by 
  \begin{equation*}
    L_-\beta = -\Delta\beta + 2 S(\nabla\beta) + \beta\big(\div S - 
    \tfrac{1}{2}|\two|^2 - |S|^2 - \Psi_- \big) \,,
  \end{equation*}
  and $\Psi_- = \tfrac{1}{4}\RiemLpmmp - \tfrac{1}{2}\RicLpm$.
\end{corollary}
If we assume that $\Sigma\subset M$, where $M$ is a three dimensional
spacelike surface, then $\Sigma$ can be deformed in the direction of
$\nu$, the normal of $\Sigma$ in $M$. The deformation of $\theta^+$
then turns out to be the following.
\begin{corollary}
  \label{cor:LM}
  Assume $\Sigma$ is a marginally trapped surface, then the
  deformation of $\theta^+$ in the spatial direction of
  $\nu:=\frac{1}{2}(l^+-l^-)$ is given by 
  \begin{equation*}
    \delta_{f\nu,w} = L_M f\,,
  \end{equation*}
  where the operator $L_M$ is given by 
  \begin{equation*}
    L_M f = -\Delta f + 2S(\nabla f)
    + f\big(\div S -|\chi|^2 + \la
    K^\Sigma,\chi\ra - |S|^2 - \Psi_M \big)\,,
  \end{equation*}
  and $\Psi_M=\tfrac{1}{4}\RiemLpmmp + \RicL(\nu,l^+)$.
\end{corollary}
\begin{remark}
  \begin{mynum}
  \item
    \label{rem:rewrite_psi}
    Using the Gauss equation \eqref{eq:gauss_tr}, we can rewrite the
    expression for $L_M$ as follows
    \begin{equation}
      \label{eq:LM_alt}
      L_M f = -\Delta f + 2 S(\nabla f) + f\big(\div S
      -\tfrac{1}{2}|\chi|^2 - |S|^2 + \tfrac{1}{2}\ScalSig - \tilde\Psi_M\big)\,.
    \end{equation}
    Here $\tilde\Psi_M= G(n,l^+)$ where $G = \RicL -
    \tfrac{1}{2}\ScalL\,h$  denotes the Einstein tensor of $h$.
    
    Note that in view of the Gauss and Codazzi equations of the embedding
    $M\hookrightarrow L$, equations \eqref{eq:gauss_ML_tr} and
    \eqref{eq:codazzi_ML_tr}, the term $\tilde\Psi_M$ can be rewritten as
    \begin{equation}
      \label{eq:psi_M_tilde}
      \tilde\Psi_M 
      =  
      \tfrac{1}{2}\big(\ScalM + (\tr K)^2 - |K|^2\big)
      - \la \divM K - \nabM \tr  K, \nu \ra      
      =: 8\pi\big(\mu-J(\nu)\big)\,,
    \end{equation}
    where $8\pi J = \divM K - \nabM \tr K$ is the projection of
    $G(n,\cdot)$ to $M$ and $16\pi \mu = \ScalM + (\tr K)^2 - |K|^2 =
    G(n,n)$. The dominant energy condition is equivalent to $|J|\leq
    \mu$.  Thus, if the dominant energy condition holds,
    $\tilde\Psi_M$ turns out to be non-negative.
  \item
    The same procedure gives that we can write $L_-$ as 
    \begin{equation}
      L_- f = -\Delta f + 2 S(\nabla f) + f\big(\div S
      - |S|^2 + \tfrac{1}{2}\ScalSig - \tilde\Psi_-\big)\,.      
    \end{equation}
    with $\tilde\Psi_-=G(l^+,l^-)$. Note that $\tilde\Psi_-$ is
    non-negative if the dominant energy condition holds. However, this
    representation does not contain a term $|\chi|^2$, which does not
    allow us to get estimates on $\sup |\chi|^2$. However, in the case
    of strict $L_-$ stability there is a sheet $M$ such that the
    surface is $L_M$-stable. We can then apply the subsequent results
    to get the estimates of theorem \ref{thm:main-intro} in this case.
    \fillbox
  \end{mynum}
\end{remark}

%% file: stable.tex
\section{Stability of marginally outer trapped surfaces}
\label{sec:stable}
As before, consider a four dimensional space time $L^4$, with a three
dimensional spacelike slice $M^3$. As in the previous sections, the
future directed unit normal to $M$ in $L$ will be denoted by $n$. In
$M$ consider a two dimensional surface $\Sigma$, such that there
exists a global unit normal vector field $\nu$ of $\Sigma$ in $M$. The
vector fields $n$ and $\nu$ span the normal bundle of $\Sigma$ in $L$
and give rise to two canonical null vectors $l^\pm =n\pm\nu$.  Again
we use the shorthand $\chi=\chi^+$.

In this section we will introduce two notions of stability for a
marginally trapped surface. These are related to variations of the
surface in different directions. The first definition is equivalent to
definition 2 in \cite{Andersson-Mars-Simon:2005}. There a \emph{stably
  outermost marginally outer trapped surface}, is defined as surface,
on which the principal eigenvalue of $L_M$ is positive. Here an
\emph{$L_M$-stable MOTS} is defined as follows.
\begin{definition}
  A two dimensional surface $\Sigma\subset M\subset L$ is called a
  \emph{$L_M$-stable marginally outer trapped surface} if
  \begin{mynum}
  \item $\Sigma$ is marginally trapped with respect to $l^+$ , that is $\theta^+=0$.
  \item There exists a function $f\geq 0, f\not\equiv 0$ such that $L_M f\geq
    0$\,. Here $L_M$ is the operator from corollary \ref{cor:LM}.
  \end{mynum}
\end{definition}
\begin{remark}
  \label{rem:lm_stable}
  \begin{mynum}
  \item Although $L_M$ is not formally self-adjoint, the eigenvalue of
    $L_M$ with the smallest real part is real and non-negative (cf.
    \cite[Lemma 1]{Andersson-Mars-Simon:2005}). This definition is
    equivalent to saying, that the principal eigenvalue of $L_M$ is
    nonnegative. This is seen as follows:
    
    Let $\lambda$ be the principal eigenvalue $L_M$. Then, since
    $\lambda$ is real, the
    $L^2$-adjoint $L_M^*$ of $L_M$ has the same principal eigenvalue and a
    corresponding eigenfunction $g>0$. Pick $f\geq 0$ as in the definition
    of $L_M$-stability, ie. $L_M f\geq 0$. Then compute
    \begin{equation*}
      \lambda \int_\Sigma f g\dmu = \int_\Sigma f L_M^* g \dmu =
      \int_\Sigma L_M f g \dmu \,.
    \end{equation*}
    As $f\geq 0$, $f\not\equiv 0$, $g>0$ and $L_M f\geq 0$, this implies
    $\lambda\geq 0$.
    
    The eigenfunction $\psi$ of $L_M$ with respect to the principal eigenvalue
    does not change sign. Therefore it can be chosen positive,
    $\psi>0$. Thus, the definition in fact is equivalent to the
    existence of $\psi>0$ such that $L_M\psi = \lambda\psi\geq 0$. We will
    use this fact frequently in the subsequent sections.
    Note that $L_M$-stability is equivalent to the notion of a
    \emph{stably outermost MOTS} in \cite[Definition
      2]{Andersson-Mars-Simon:2005}.  
  \item
    The conditions from the above definition are satisfied in the
    following situation. Let $\Sigma=\del\Omega$ be the boundary of
    the domain $\Omega$ and satisfy $\theta^+=0$. Furthermore assume
    that there is a neighborhood $U$ of $\Sigma$ such that the
    exterior part $U\setminus \Omega$ does not contain any trapped
    surface, ie. a surface with $\theta^+< 0$. Then $\Sigma$ is
    stable. Assume not. Then the principal eigenvalue would be
    negative and the corresponding eigenfunction $\psi$ would satisfy
    $L_M\psi<0$, $\psi>0$. This would imply the existence of trapped surfaces
    outside of $\Sigma$, since the variation of $\Sigma$ in direction
    $\psi\nu$ would decrease $\theta^+$.
  \fillbox
  \end{mynum}
\end{remark}
Note that the condition $\theta^+=0$ does not depend on the
choice of the particular frame. Therefore, to say that a surface is
marginally trapped, we do not need any additional information. In
contrast the notion of stability required here does depend on the
frame, since clearly there is no distinct selection of $\nu$ when
only $\Sigma$ --- and not $M$ --- is specified.  

To address this issue, we introduce the second notion of stability of
marginally outer trapped surfaces, namely with reference to the
direction $-l^-$. This definition is more in spirit of Newman
\cite{Newman:1987} and recent interest in the so called dynamical
horizons \cite{Ashtekar-Krishnan:2003, Ashtekar-Galloway:2005}.
\begin{definition}
  A two dimensional surface $\Sigma\subset M\subset L$ is called a
  \emph{$L_-$-stable marginal outer trapped surface ($L^-$-stable MOTS)} if 
  \begin{mynum}
  \item $\Sigma$ is marginally trapped with respect to $l^+$ , that is $\theta^+=0$.
  \item There exists a function $f\geq 0, f\not\equiv 0$ such that
    $L_- f\geq 0$\,. Here $L_-$ is the operator from corollary \ref{cor:Lminus}.
  \end{mynum}
\end{definition}
\begin{remark}
  It turns out that this notion of stability does not depend on the
  choice of the null frame. This is due to the natural transformation
  law of the stability operator $L_-$ when changing the frame
  according to $\tilde l^+ = f l^+$ and $\tilde l^{-} = f^{-1}
  l^-$. Then the operator $\tilde L_-$ with respect to this frame
  satisfies $f^{-1}\tilde L (f\beta) = L\beta$ for all functions
  $\beta\in C^\infty(\Sigma)$, as it is expected from the facts that
  $\tilde \theta^+ = f \theta^+$ and $-\beta l^- = -\beta f \tilde
  l^-$.  \fillbox
\end{remark}
\begin{remark}
  \begin{mynum}
  \item
    Remark \ref{rem:lm_stable} is also valid here, in particular the
    definition implies that there exists a function $\psi>0$ with
    $L_-\psi\geq 0$.
  \item
    Technically speaking, the equation for a marginally trapped
    surface prescribes the mean curvature H of $\Sigma$ in $M$ to
    equal minus the value of a function $P : TM \to \IR : (p,v)\mapsto
    tr K -K_{ij}\nu^i\nu^j$, namely $H(p)=-P(p,\nu)$ for all
    $p\in\Sigma$. This is a degenerate quasilinear elliptic equation
    for the position of the surface. These equations do not allow
    estimates for second derivatives without any additional
    information. This is where the two stability conditions come into
    play. They give the additional piece of information needed in the
    estimates as in the case for stable minimal surfaces.
    \fillbox
  \end{mynum}
\end{remark}
The two notions of stability above imply the positivity of certain
symmetric differential operators as it was noticed in
\cite{Galloway-Schoen:2005} for the operator $L_M$. However, the
inequality there is not quite sufficient for our purposes, it needs
some further rearrangement. This is the content of the following
Lemmas. Basically lemmas \ref{lemma:LM_sym_stab} and
\ref{lemma:LM_sym_stab} are the only way how stability is used in the
subsequent estimates. Basically one could use these, in paricular
equation \eqref{eq:LM_sym_stab_GS} to define a notion of symmetrized
stability for MOTS.
\begin{lemma}
  \label{lemma:LM_sym_stab}
  If $\Sigma$ is a stable MOTS, then for all $\eps>0$ for all $\eta\in
  C_c^\infty(\Sigma)$ the following inequality holds
  \begin{equation*}
    \int_\Sigma
    \eta^2 |\chi|^2 \dmu
    \leq
    (1+\eps) \int_\Sigma
    |\nabla\eta|^2 + \eta^2\big((4\eps)^{-1}|K^\Sigma|^2 - \Psi_M\big)\dmu.
  \end{equation*}
\end{lemma}
\begin{proof}
  Take $f$ as in the definition of a stable MOTS. From remark
  \ref{rem:lm_stable} we can assume $f>0$. Then $f^{-1}L_M f\geq 0$.
  Multiply this relation by $\eta^2$, integrate, and expand $L_M$ as
  in corollary \ref{cor:LM}. This yields
  \begin{equation*}
    0 \leq\int_\Sigma  \eta^2\big( - f^{-1}\Delta f + 2f^{-1} S(\nabla f)
    + \div S -|\chi|^2 + \la K^\Sigma,\chi\ra  - |S|^2 - \Psi_M\big) \dmu\,.
  \end{equation*}
  By sorting terms, and partially integrating the Laplacian and the
  divergence term, we obtain
  \begin{equation*}
    \begin{split}
      &
      \int_\Sigma \eta^2 |\chi|^2 + \eta^2 \big( f^{-2}|\nabla f|^2 - 2 f^{-1}
      S(\nabla f) + |S|^2 \big) \dmu
      \\
      &\quad
      \leq
      \int_\Sigma 2\eta\la\nabla\eta,f^{-1}\nabla f - S \ra +
      \eta^2|\chi|\,|K^\Sigma| -\eta^2 \Psi_M\dmu\,.
    \end{split}
  \end{equation*}
  By the Schwarz inequality
  \begin{equation*}
    \int_\Sigma 2\eta\la\nabla\eta,f^{-1}\nabla f - S \ra\dmu
    \leq \int_\Sigma |\nabla \eta|^2 + \eta^2 |f^{-1}\nabla f - S|^2\dmu
  \end{equation*}
  and for any $\eps>0$
  \begin{equation*}
    \int_\Sigma \eta^2|K^\Sigma|\,|\chi| \dmu \leq (4\eps)^{-1}\int_\Sigma
    \eta^2|K^\Sigma|^2 \dmu + \eps\int_\Sigma
    \eta^2|\chi|^2\,.
  \end{equation*}    
  Cancelling the terms $\int_\Sigma \eta^2|f^{-1}\nabla f - S|^2\dmu$
  and $\eps\int_\Sigma \eta^2|\chi|^2\dmu$ on both sides, we conclude
  the claimed inequality.
\end{proof}
The following lemma is based on the original computation of \cite{Galloway-Schoen:2005}.
\begin{lemma}
  \label{lemma:LM_sym_stab_alt}
  If $\Sigma$ is a stable MOTS, then for all $\eps>0$, there exists
  $C(\eps^{-1})$ such that for all $\eta\in C^\infty_c(\Sigma)$ the
  following inequality holds:
  \begin{equation*}
    \int_\Sigma \eta^2 |\chi|^2 \dmu \leq (1+\eps)\int_\Sigma |\nabla\eta|^2
    + \eta^2 \big(|\RicM| - \tilde\Psi_M + C(\eps^{-1})|K^\Sigma|^2\big)\dmu.  
  \end{equation*}
\end{lemma}
\begin{proof}
  We proceed as in the computation of lemma~\ref{lemma:LM_sym_stab},
  but with the alternative representation~\eqref{eq:LM_alt} for $L_M$.
  As in \cite{Galloway-Schoen:2005}, we get
  \begin{equation}
    \label{eq:LM_sym_stab_GS}
    \int_\Sigma \eta^2 |\chi|^2 \dmu \leq \int_\Sigma 2|\nabla\eta|^2
    + \eta^2 \big(\ScalSig - 2\tilde\Psi_M\big)\dmu.  
  \end{equation}
  We use Gauss' equation to replace $\ScalSig$ in the following way
  \begin{equation*}
    \ScalSig=\ScalM -2\RicM(\nu,\nu) + H^2 - |A|^2,
  \end{equation*}
  where $A$ is the second fundamental form of $\Sigma\subset M$ and
  $H$ is the mean curvature.
  We can move the $|A|^2$ term to the left hand side. Then $H^2 = P^2$ by
  $\theta^+=0$ and thus $H^2 \leq 2|K^\Sigma|^2$. The remaining terms
  are controlled by $|\RicM|$. Inserting this, we find that 
  \begin{equation*}
    \int_\Sigma \eta^2 (|\chi|^2 + |A|^2) \dmu \leq 2\int_\Sigma |\nabla\eta|^2
    + \eta^2 \big(|\RicM| + |K^\Sigma|^2 - \tilde\Psi_M\big)\dmu.  
  \end{equation*}
  Now fix $\eps>0$. Since $\chi = A + K^\Sigma$ we can estimate
  \begin{equation*}
    \begin{split}
      2|\chi|^2
      &\leq
      (1+\eps)|\chi|^2 + (1-\eps)(|A|^2 + 2\la A,
      K^\Sigma\ra + |K^\Sigma|^2)
      \\
      &\leq
      (1+\eps)(|\chi|^2 + |A|^2)
      + (2-2\eps +(2\eps)^{-1})|K^\Sigma|^2.
    \end{split}
  \end{equation*}
  Inserting this into the above inequality we find the claimed inequality.
\end{proof}
A similar, but fundamentally different inequality holds in the case of
$L_-$-stability. The fundamental difference is that the gradient term
on the right has a factor of a little more than two, instead of a
little more than one, as with the $L_M$-stability. In view of lemma
\ref{lemma:LMstab_to_Lminus_stab} this factor of two is not at all
surprising. This factor is the reason why the procedure in
section~\ref{sec:apriori} does not work to give curvature estimates
for $L_-$-stable surfaces.
\begin{lemma}
  \label{lemma:Lminus_sym_stab}
  If $\Sigma$ is a $L_-$-stable MOTS, then for all $\eps>0$ the
  following inequality holds:
  \begin{equation*}
    \int_\Sigma |\chi|^2 \eta^2\dmu \leq 2(1+\eps)\int_\Sigma
    |\nabla\eta|^2 + \eta^2 \big((2\eps)^{-1}|K^\Sigma|^2 - \Psi_-\big)\dmu.
  \end{equation*}
\end{lemma}
We conclude with the remark that $L_M$-stability implies
$L_-$-stability.
\begin{lemma}
  \label{lemma:LMstab_to_Lminus_stab}
  Let $(L,h)$ satisfy the null energy condition, i.e. assume that for
  all null vectors $l$ we have that $\RicL(l,l)\geq 0$. Then if
  $\Sigma$ is an $L_M$-stable MOTS, then it is also $L_-$-stable.
\end{lemma}
\begin{proof}
  We use the notation from section \ref{sec:linearization}, where we
  introduced the linearization of $\theta^+$. For any function $f$ compute 
  \begin{equation*}
    L_M f - L_- f 
    = \delta_{f\nu,w}\theta^+ - \tfrac{1}{2}\delta_{fl^-,w}\theta^+ 
    = \tfrac{1}{2}\delta_{f l^+,w} \theta^+
    = -\tfrac{1}{2} f \big(|\chi|^2 + \RicL(l^+,l^+)\big)\,.
  \end{equation*}
  If $f>0$, then by the null energy condition, the right hand side is
  non-positive. If in addition $L_M f\geq 0$, as in the definition of
  $L_M$-stability, then this implies that
  \begin{equation*}
    L_- f \geq L_M f \geq 0\,.
  \end{equation*}
  Hence $\Sigma$ is also $L_M$ stable.
\end{proof}


%% file: apriori.tex
\section{A priori estimates}
\label{sec:apriori}
In this section we derive the actual estimates for stable outermost
marginally trapped surfaces. All but the most basic estimates hold
only for $L_M$-stable surfaces, as defined in
section~\ref{sec:stable}. This is due to the factor of two appearing
in front of the gradient term in lemma~\ref{lemma:Lminus_sym_stab},
which does not allow us to carefully balance the Simons inequality and
the stability inequalities.

Throughout this section we will make the assumption, that the surfaces
in question have \emph{locally uniformly finite area}.
\begin{definition}
  If there exists $r>0$ and $a<\infty$, such that for all $x\in
  \Sigma$
  \begin{equation}
    \label{eq:lfa}
    | B_\Sigma (x,r) | \leq a,
  \end{equation}
  then we say that $\Sigma$ has $(r,a)$-locally uniformly finite area.
\end{definition}
Here $B_\Sigma (x,r)$ denotes the the ball of radius $r$ around $x$ in
$\Sigma$.  In the sequel we will denote $B_\Sigma(x,r)$ by $B(x,r)$.
The estimates below work in exactly the same way if intrinsic balls
are replaced by extrinsic balls.  Later, we will derive such bounds
for stable MOTS in terms of the ambient geometry.

We first begin with the observation that the stability of a MOTS gives
a local $L^2$-estimate for the shear tensor $\chi=\chi^+$.

In the sequel, for a tensor $T$, we denote by $\|T\|_\infty
=\sup_\Sigma|T|$. That is, $\infty$-norms are taken on $\Sigma$
only. Constant are always denoted by C, if we want to clarify the
dependence of the constants, we denote by $C(a,b,\ldots)$ a constant
that depends on the quantities $a,b\ldots$ in  such a way that $C$
deteriorates as $a + b + \ldots \to\infty$. 
\begin{lemma}
  \label{lemma:sigma_l2}
  Suppose $\Sigma$ is an $L_M$-stable MOTS with $(r,a)$-locally
  uniformly finite area. Then for all $x\in \Sigma$
  \begin{equation*}
    \int_{B(x,r/2)} |\chi|^2 \dmu \leq C(r^{-1},a,\|K^\Sigma\|_{\infty},\|\Psi_M\|_{\infty})\,.
  \end{equation*}
  Alternatively, the constant can be chosen to depend on
  $\|\tilde\Psi_M\|_{\infty}$ and $\|\RicM\|_\infty$ instead of
  $\|\Psi_M\|_{\infty}$.
\end{lemma}
\begin{proof}
  The desired bound is easily derived from lemma
  \ref{lemma:LM_sym_stab} or \ref{lemma:LM_sym_stab_alt}. We will
  restrict ourselves to the proof of the first statement. To this end,
  fix $\eps = \frac{1}{2}$, $x\in\Sigma$ and choose a cut-off function
  $\eta\geq 0$ such that $\eta\equiv 1$ on $B(x,r/2)$, $\eta = 0$ on
  $\del B(x,r)$, and $|\nabla \eta| \leq 4r^{-1}$. The left hand side
  of the equation in lemma \ref{lemma:LM_sym_stab} is then is an upper
  bound for the left hand side in the claim, whereas the right hand
  side can be estimated by the claimed quantities.
\end{proof}
This estimate can also be derived from $L_-$-stability:
\begin{lemma}
  \label{lemma:sigma_l2_Lminus}
  Suppose $\Sigma$ is an $L_-$-stable MOTS with $(r,a)$-locally
  uniformly finite area. Then for all $x\in \Sigma$
  \begin{equation*}
    \int_{B(x,r/2)} |\chi|^2 \dmu \leq C(r^{-1},a,\|K^\Sigma\|_{\infty},\|\Psi_-\|_{\infty})\,.
  \end{equation*}
\end{lemma}
\begin{proposition}
  \label{prop:sigma_lp}
  Let $\Sigma$ be an $L_M$-stable MOTS. For any $\eps>0$, any
  $p\geq 2$, and any function $\eta$ we have the estimate
  \begin{equation*}   
    \begin{split}
      \int_\Sigma \eta^2 |\chi|^{p+2}\dmu      
      &\leq
      \tfrac{p^2}{4}(1+\eps) \int_\Sigma
      \eta^2 |\chi|^{p-2}\big|\nabla |\chi|\big|^2\dmu
      \\
      &\quad+ C(\eps^{-1},\|\Psi_M\|_\infty, \|K^\Sigma\|_\infty)
      \int_\Sigma\big(\eta^2 + |\nabla\eta|^2\big)|\chi|^p\dmu.
    \end{split}
  \end{equation*}
  Alternatively, we can make the constant on the right hand side
  to be of the form $C(\eps^{-1},\|\tilde\Psi_M\|_\infty,
  \|K^\Sigma\|_\infty,\|\RicM\|_\infty))$.
\end{proposition}
This proposition also holds for $L_-$-stable surfaces with appropriate
modifications of the dependencies of $C$, and a general factor of two
on the right hand side. This factor of is the reason, why an argument
like the subsequent one fails to give curvature estimates for
$L_-$-stable MOTS.
\begin{proof}
  We will restrict to the proof of the first statement, since the
  other is proved similarly. From lemma \ref{lemma:LM_sym_stab} we
  find that for any $\delta>0$ there is $C(\delta^{-1})$ such that for
  all functions $\phi$, we have
  \begin{equation*}
    \int_\Sigma \phi^2 |\chi|^2 \dmu \leq (1+\delta) \int_\Sigma
    |\nabla\phi|^2 + \phi^2\big(C(\delta^{-1})|K^\Sigma|^2 - \Psi_M\big)\dmu.
  \end{equation*}
  We substitute $\phi$ by $\eta|\chi|^{p/2}$. To this end compute 
  \begin{equation*}
    \nabla (\eta|\chi|^{p/2}) = \nabla\eta|\chi|^{p/2} + \eta\,\tfrac{p}{2}|\chi|^{p/2-1}\nabla|\chi|.
  \end{equation*}
  For any $\delta>0$ we thus can estimate
  \begin{equation*}
    |\nabla (\eta|\chi|^{p/2})|^2 \leq (1-\delta)
    \frac{p^2}{4}|\chi|^{p-2} \eta^2\big|\nabla|\chi|\big|^2 + C(\delta^{-1})|\nabla\eta|^2|\chi|^p.
  \end{equation*}
  Inserting this estimate into the first inequality, we find
  \begin{equation*}
    \begin{split}
      \int_\Sigma \eta^2 |\chi|^{p+2} \dmu
      &\leq
      (1+\delta)^2\,\tfrac{p^2}{4} \int_\Sigma
      \eta^2|\chi|^{p-2} \big|\nabla|\chi|\big|^2
      \\
      &\quad
      + \int_\Sigma C(\delta^{-1})|\nabla\eta|^2 + \eta^2|\chi|^p\big(C(\delta^{-1})|K^\Sigma|^2 - \Psi_M\big)\dmu.
    \end{split}
  \end{equation*}
  Adjusting $\delta$ yields the claimed estimate.
\end{proof}
We now aim for an estimate on the gradient term on the right hand side
of the estimate in proposition \ref{prop:sigma_lp}. The main tool will
be the Simons identity from section \ref{sec:simons}. To avoid that the
estimated depend on derivatives of curvature, we use similar
techniques as in \cite{Metzger:2004}.
\begin{proposition}
  \label{prop:sigma_grad}
  Let $\Sigma$ be an $L_M$-stable MOTS. Then there exists $p_0>2$ such
  that for $2\leq p \leq p_0$ and all functions $\eta$ we have the estimate
  \begin{equation*}  
    \begin{split}
      &\int_\Sigma \eta^2|\chi|^{p-2} \big|\nabla|\chi|\big|^2 \dmu
      \\
      &\leq
      C(p, \|\Psi_M\|_\infty, \|K^\Sigma\|_\infty, \|Q\|_\infty, \|\RiemL^\Sigma\|_\infty,
      \|S\|_\infty)\int_\Sigma (\eta^2 + |\nabla\eta|^2)|\chi|^p +
      \eta^2 |\chi|^{p-2}.
    \end{split}
  \end{equation*}
  Alternatively, as before, we can replace the dependence of the
  constant on $\|\Psi_M\|_\infty$ by a dependence on
  $\|\tilde\Psi_M\|_\infty$ and $\|\RicM\|_\infty$.
\end{proposition}
Before we can start the proof of the proposition, we state the
following lemma. It states an improved Kato's inequality similar to
\cite{Schoen-Yau:1981}. A general reference for such inequalities is
\cite{Calderbank-Gauduchon-Herzlich:2000}.
\begin{lemma}
  \label{lemma:sy_trick}
  On a surface $\Sigma$ with $\theta^+=0$ we have the estimate
  \begin{equation*}
    |\nabla\chi|^2 - \big|\nabla|\chi|\big|^2 \geq \tfrac{1}{33}
     \big( \big|\nabla|\chi|\big|^2 + |\nabla\chi|^2\big) -
     c\,\big(|Q|^2 + |S|^2|\chi|^2\big)\,.
  \end{equation*}  
  Here $c$ is a purely numerical constant.
\end{lemma}
\begin{proof}
  The proof goes along the lines of a similar argument in Schoen and
  Yau in \cite[p. 237]{Schoen-Yau:1981}, but for the sake of completeness, we
  include a sketch of it here.
  
  In the following computation we do not use the Einstein summation
  convention and work in a local orthonormal frame for $T\Sigma$. Let
  $T := |\nabla \chi|^2 - \big|\nabla|\chi|\big|^2$. We compute
  \begin{equation*}
    \begin{split}
      |\chi|^2 T 
      &= |\chi|^2|\nabla\chi|^2 - \tfrac{1}{4}
      \big|\nabla|\chi|^2\big|^2
      \\
      &= \sum_{i,j,k,l,m} (\chi_{ij}\nabla_k\chi_{lm})^2 -
      \sum_k\Big(\sum_{ij}\chi_{ij}\nabla_k\chi_{ij}\Big)^2
      \\
      &= \tfrac{1}{2}\sum_{i,j,k,l,m}\Big(\chi_{ij}\nabla_k\chi_{lm}
      -\chi_{lm}\nabla_k\chi_{ij}\Big)^2\,.
    \end{split}
  \end{equation*}
  In the last term consider only summands with $i=k$ and $j=m$. This
  gives
  \begin{equation*}
      |\chi|^2 T 
      \geq \tfrac{1}{2} \sum_{i,j,l} \Big(
      \chi_{ij}\nabla_i\chi_{jl} -
      \chi_{jl}\nabla_i\chi_{ij}\Big)^2
      \geq \tfrac{1}{8}\sum_{l}\Big(\sum_{i,j}\chi_{ij}\nabla_i\chi_{jl}
      - \chi_{jl}\nabla_i\chi_{ij}\Big)^2\,.
  \end{equation*}
  Use the Codazzi equation \eqref{eq:codazzi_coord} to swap indices
  in the gradient terms. We arrive at
  \begin{equation*}
      |\chi|^2 T 
      \geq \tfrac{1}{8} \sum_{l} \Big( \sum_{i,j} \big(\chi_{ij}\nabla_l
      \chi_{ij} + \chi_{ij}Q_{ilj} - \chi_{lj}Q_ {iji}\big)
      +\sum_i \big(\theta S_i\chi_{il} -
      \chi_{il}\nabla_i\theta\big) - |\chi|^2 S_l \Big)^2\,.
  \end{equation*}
  By the fact that $(a-b)^2 \geq \frac{1}{2}a^2 - b^2$, this implies
  \begin{equation*}
     \begin{split}
       |\chi|^2 T 
       &\geq \tfrac{1}{16} \sum_l \Big(\sum_{i,j}
       \chi_{ij}\nabla_l\chi_{ij}\Big)^2
       \\
       &\phantom{\geq}
       - \tfrac{1}{8} \sum_l\Big(\sum_{i,j}
       \big(\chi_{ij}Q_{ilj}-\chi_{lj}Q_{iji}\big) +
       \sum_i\chi_{il}(\theta S_i - \nabla_i\theta) -
       |\chi|^2S_l\Big)^2
       \\
       &\geq 
       \tfrac{1}{16} |\chi|^2\big|\nabla|\chi|\big|^2 - c\,
       |\chi|^2\big(|Q|^2 + |S|^2|\chi|^2\big)\,.
     \end{split}
  \end{equation*}
  Dividing by $|\chi|^2$, we get
  \begin{equation*}
    |\nabla\chi|^2 - \big|\nabla|\chi|\big|^2 \geq
    \tfrac{1}{16}\big|\nabla|\chi|\big|^2 -c \big(|Q|^2 +
    |S|^2|\chi|^2\big)\,.
  \end{equation*}
  Adding $\frac{1}{32} \big(|\nabla\chi|^2 - \big|\nabla|\chi|\big|^2\big)$
  to both sides of this inequality and multiplying by $\frac{32}{33}$ yields
  the desired estimate. 
\end{proof}
\begin{proof}[Proof of proposition \ref{prop:sigma_grad}.]
  We will restrict to the proof of the first statement, since the
  other is proved similarly.
  Compute
  \begin{equation*}
    \Delta |\chi|^2 = 2 |\chi|\Delta|\chi| + 2
    \big|\nabla|\chi|\big|^2\,.
  \end{equation*}
  On the other hand
  \begin{equation*}
    \Delta |\chi|^2 = 2 \chi_{ij}\Delta\chi_{ij} + 2|\nabla
    \chi|^2\,.
  \end{equation*}
  Subtracting these equations yields
  \begin{equation*}
    |\chi|\Delta|\chi| = \chi_{ij}\Delta \chi_{ij} + |\nabla
    \chi|^2 - \big|\nabla|\chi|\big|^2\,.
  \end{equation*}
  In the case $\theta^+=0$, the Simons identity from lemma
  \ref{lemma:simons} gives 
  \begin{equation*}
    \begin{split}
      \chi_{ij}\lap\chi_{ij} &= \chi_{ij}\big(\RiemL_{kilk}\chi_{lj} +
      \RiemL_{kilj}\chi_{kl}\big) - |\two|^2 |\chi|^2
      - P\chi_{ij}\chi_{jk}\chi_{ki}
      \\ & \phantom{=}
      + \chi_{ij} \nabla_k \big(Q_{kij} - \chi_{kj}S_i +
      \chi_{ij} S_k\big)
      + \chi_{ij} \nabla_i \big(Q_{kjk} + \chi_{jk} S_k\big)\,.
    \end{split}
  \end{equation*}
  Note that $\chi_{ij}\chi_{jk}\chi_{ki} = \tr
  (\chi^3)$, and the trace of a $2\times 2$ matrix $A$ satisfies the
  relation $\tr A^3 = \tr A (\tr A^2 - \det A)$. Since $\chi$ is
  traceless, this term vanishes.  In addition
  $|\two|^2=\la\chi,\chi^-\ra = |\chi|^2 - 2\la
  K^\Sigma,\chi\ra$. 

  As we are not interested in the particular form of some terms, to
  simplify notation, we introduce the $*$-notation. For two tensors
  $T_1$ and $T_2$, the expression $T_1 * T_2$ denotes linear
  combinations of contractions of $T_1 \tensor T_2$.

  To remember that in the above equation we need to evaluate $\RiemL$
  only on vectors tangential to $\Sigma$, we use the projection of
  $\RiemL$ to $T\Sigma$ and denote this by $\RiemL^\Sigma$. Then the
  above equations combine to
  \begin{equation}
    \label{eq:simons_final}
      - |\chi|\Delta|\chi| +
       |\nabla\chi|^2 -
       \big|\nabla|\chi|\big|^2       
       =|\chi|^4 + |\chi|^2 * \chi * K^\Sigma 
       + \chi * \chi * \RiemL^\Sigma
      + \chi * \nabla \big(Q + \chi * S\big)\,.
  \end{equation}
  Multiply this  equation by $\eta^2|\chi|^{p-2}$ and
  integrate. This yields
  \begin{equation*}
    \begin{split}
      &\int_\Sigma - \eta^2|\chi|^{p-1} \Delta|\chi| + \eta^2|\chi|^{p-2}\big(
      |\nabla \chi|^2 - \big|\nabla|\chi|\big|^2\big) \dmu
      \\
      &=\!
      \int_\Sigma \eta^2\big(|\chi|^{p+2}\!+ |\chi|^p \chi * K^\Sigma\! +
      |\chi|^{p-2} \chi * \chi * \RiemL^\Sigma + |\chi|^{p-2} \chi *
      \nabla(Q+ \chi * S)\big) \dmu.
    \end{split}
  \end{equation*}
  Next, do a partial integration on the term including the Laplacian
  and on the last term on the second line. We find that
  \begin{equation}
    \label{eq:gradest_1}
    \begin{split}
      &\int_\Sigma \eta^2 (p-1) |\chi|^{p-2} \big|\nabla|\chi|\big|^2 +
      \eta^2 |\chi|^{p-2}\big(|\nabla \chi|^2 - \big|\nabla|\chi|\big|^2\big)
      \dmu
      \\
      &\leq
      \int_\Sigma \eta^2|\chi|^{p+2} \dmu + c\int_\Sigma
      \eta|\chi|^{p-1}|\nabla\eta|\,|\nabla|\chi|| + \eta|\chi|^{p-1}|\nabla\eta|(|Q|+|\chi||S|)\dmu 
      \\
      &\
      + c\int_\Sigma \eta^2 \big\{|\chi|^{p+1}
      |K^\Sigma| + |\chi|^p|\RiemL^\Sigma| +
      |\chi|^{p-2}\big(|\nabla\chi|+ \big|\nabla|\chi|\big|\big)(|Q| +
      |\chi||S|)\big\} \dmu\,.
    \end{split}
  \end{equation}
  Here $c$ is a purely numerical constant. For any $\eps>0$, we can estimate 
  \begin{equation*}
    c\int_\Sigma \eta^2|\chi|^{p+1} |K^\Sigma|\dmu
    \leq
    \eps\int_\Sigma \eta^2|\chi|^{p+2}\dmu + C(\eps^{-1}) \int_\Sigma
    \eta^2|\chi|^p |K^\Sigma|^2\dmu
  \end{equation*}
  as well as 
  \begin{equation*}
    \begin{split}
      &c\int_\Sigma \eta^2|\chi|^{p-2}\big(|\nabla\chi|+ \big|\nabla|\chi|\big|\big)(|Q| +
      |\chi||S|) \dmu
      \\
      &\quad\leq 
      \eps \int_\Sigma \eta^2|\chi|^{p-2} \big( |\nabla \chi|^2 +
      \big|\nabla|\chi|\big|^2\big)\dmu + C(\eps^{-1}) \int_\Sigma
      \eta^2\big(|\chi|^p|S|^2 + |\chi|^{p-2}|Q|^2\big)\dmu\,.      
    \end{split}
  \end{equation*}
  In addition we estimate
  \begin{equation*}
    c\int_\Sigma \eta|\chi|^{p-1}|\nabla\eta|\,|\nabla|\chi||\dmu
    \leq
    \eps \int_\Sigma \eta^2\big|\nabla|\chi|\big|^2|\chi|^{p-2} +
    C(\eps^{-1})\int_\Sigma |\nabla\eta|^2|\chi|^p\dmu,
  \end{equation*}
  and
  \begin{equation*}
    \int_\Sigma \eta|\chi|^{p-1}|\nabla\eta|(|Q|+|\chi||S|)\dmu
    \leq \int_\Sigma |\nabla\eta|^2|\chi|^p +
    \eta^2\big(|\chi|^{p-2}|Q|^2 + |\chi|^p|S|^2\big)\dmu
  \end{equation*}
  Inserting these estimates into the estimate \eqref{eq:gradest_1}, gives
  \begin{equation}
    \label{eq:gradest_2}
    \begin{split}
      &\int_\Sigma \eta^2 (p-1) |\chi|^{p-2} \big|\nabla|\chi|\big|^2 +
      \eta^2|\chi|^{p-2}\big(|\nabla \chi|^2 - \big|\nabla|\chi|\big|^2\big)
      \dmu
      \\
      &\leq \int_\Sigma (1+\eps)\eta^2|\chi|^{p+2} + 2\eps \eta^2|\chi|^{p-2} \big( |\nabla\chi|^2
      +\big|\nabla|\chi|\big|^2\big)\dmu + C(\eps^{-1})\int_\Sigma |\nabla\eta|^2|\chi|^p\dmu
      \\
      &\quad + C(\eps^{-1}, \|K^\Sigma\|_\infty, \|\RiemL^\Sigma\|_\infty, \|S\|_\infty,\|Q\|_\infty)
      \int_\Sigma \eta^2\big(|\chi|^p + |\chi|^{p-2}\big)\dmu\,.
    \end{split}
  \end{equation}
  We apply lemma \ref{lemma:sy_trick} to estimate the second term 
  on the left hand side from below by $\frac{1}{33}\int_\Sigma \eta^2\big(\big|\nabla|\chi|\big|^2 + |\nabla
  \chi|^2 \big)\dmu$. In addition, use proposition \ref{prop:sigma_lp} to
  estimate the first term on the right hand side. This yields
  \begin{equation*}  
    \begin{split}
      &\int_\Sigma \eta^2(p-1)|\chi|^{p-2}\big|\nabla|\chi|\big|^2 +
      (\tfrac{1}{33}-2\eps)\eta^2|\chi|^{p-2} \big(|\nabla\chi|^2 +
      |\nabla|\chi|\big|^2\big) \dmu
      \\
      &\leq \tfrac{p^2}{4} (1+\eps)^2\int_\Sigma
      \eta^2 |\chi|^{p-2}\big|\nabla|\chi|\big|^2\dmu 
      \\
      & \ 
      + C(\eps^{-1}, \|K^\Sigma\|_\infty,\|Q\|_\infty,
      \|\RiemL^\Sigma\|_\infty,\|S\|_\infty ) \int_\Sigma
      \eta^2\big(|\chi|^p + |\chi|^{p-2}\big) + |\nabla\eta|^2|\chi|^p\dmu\,.
    \end{split}
  \end{equation*}
  Choose $p_0>2$ close enough to $2$ and $\eps$ small enough, such
  that for $2<p<p_0$ the gradient term on the right hand side can be absorbed on
  the left hand side. This gives the desired estimate.
\end{proof}
Combining propositions \ref{prop:sigma_lp} and \ref{prop:sigma_grad}
with the initial $L^2$-estimate in lemma \ref{lemma:sigma_l2} gives
the following $L^p$ estimates for $|\chi|$.
\begin{theorem}
  \label{thm:sigma_lp}
  There exists $p_0>2$ such that for all $2\leq p \leq p_0$ and all
  $L_M$-stable MOTS $\Sigma$ which have $(r,a)$-locally uniformly
  finite area, the shear $\chi$ satisfies for all $x\in\Sigma$ the
  estimates
  \begin{align}
    \label{eq:sigma_Lp}
    &\int_{B(x,r/8)} |\chi|^{p+2} \dmu 
    \leq
    C,
    \\
    \label{eq:sigma_grad_pre}
    &\int_{B(x,r/8)} |\chi|^{p-2} \big|\nabla |\chi|\big|^2\!\dmu
    \leq 
    C,
    \intertext{and}
    \label{eq:sigma_grad_L2}
    &\int_{B(x,r/8)} |\nabla\chi|^2 \dmu 
    \leq 
    C\,.
  \end{align}
  Here
  $C=C(r^{-1},a,\|\Psi_M\|_\infty,\|K^\Sigma\|_\infty,\|Q\|_\infty,\|\RiemL^\Sigma\|_\infty,\|S\|_\infty)$,
  or alternatively, we can replace the dependence on $\|\Psi_M\|_\infty$
  by $\|\tilde\Psi_M\|_\infty$ and $\|\RicM\|_\infty$.
\end{theorem}
\begin{proof}  
  Fix $x\in\Sigma$ and choose $\eta$ to be a cut-off function with
  $\eta\equiv 1$ in $B(x,r/4)$ and $\eta\equiv 0$ outside of
  $B(x,r/2)$, such that $0\leq\eta\leq 1$ and $|\nabla\eta|\leq
  4r^{-1}$. Plugging this into the estimates of \ref{prop:sigma_grad}
  and \ref{prop:sigma_lp} for $p=2$, in view of the local area bound,
  and the local $L^2$-estimate from lemma \ref{lemma:sigma_l2}, yields
  $L^4$-estimates for $|\chi|$ in $B(x,r/4)$.

  We pick $p_0$ a little smaller, than the value allowed by Proposition
  \ref{prop:sigma_grad}. Then for any $2<p\leq p_0$, proceed as
  before, but now choose a cutoff function $\bar\eta$ with
  $\bar\eta\equiv 1$ in $B(x,r/8)$ and $\bar\eta\equiv 0$ outside of
  $B(x,r/4)$, such that $0\leq\bar\eta\leq 1$ and
  $|\nabla\bar\eta|\leq 8r^{-1}$. The resulting $L^p$ and
  $L^{p-2}$-norms of $|\chi|$ on the right hand side can now be
  estimated by combinations of the $L^4$-norm of $|\chi|$ and the
  local area bound.

  To see the last estimate, note that in the proof of proposition
  \ref{prop:sigma_grad}, by appropriately choosing $\eps$, we can
  retain a small portion of the term $\int_\Sigma
  \eta^2 |\chi|^{p-2}|\nabla\chi|^2\dmu$ on the right hand side.
\end{proof}  
For the next step -- the derivation of $\sup$-bounds on $\chi$ -- we
use the Hoffman-Spruck Sobolev inequality in the following form
\cite{Hoffman-Spruck:1974}.
\begin{lemma}
  \label{lemma:hs_sobolev}
  For $(M,g)$ exist constants $c^S_0,c^S_1$, such that for all
  hypersurfaces $\Sigma\subset M$ and all functions
  $f\in C^\infty(\Sigma)$ with $|\supp f|\leq c^S_0$ the following
  estimate holds:
  \begin{equation*}
    \left(\int_\Sigma |f|^2 \dmu\right)^{1/2} \leq c^S_1 \int_\Sigma |\nabla f| +
    |f H| \dmu\,.
  \end{equation*}
  Here $H$ is the mean curvature of $\Sigma$ and the constants $c^S_0,
  c^S_1$ depend only on a lower bound for the injectivity radius and
  an upper bound for the curvature of $(M,g)$.
\end{lemma}
\begin{remark}
  \label{rem:hs_sobolev}
  Replacing $f$ by $f^p$ in the above inequality and using H\"olders
  inequality gives that for all
  $1<p<\infty$ and all $f$ with $|\supp f|\leq c^S_0$ 
  \begin{equation*}
    \left(\int_\Sigma f^p\dmu\right)^{2/p} \leq c^S_p |\supp
    f|^{2/p} \int_\Sigma |\nabla f|^2 + |Hf|^2 \dmu\,.
  \end{equation*}
  The constant $c^S_p$ only depends on $c^S_1$ and $p$.
\end{remark}
\begin{theorem}
  \label{thm:main}
  Let $\Sigma$ be an $L_M$-stable MOTS with $(r,a)$-locally finite area. Then
  the shear $\chi$ satisfies the estimate
  \begin{equation*}  
    \sup_\Sigma |\chi| \leq  C.
  \end{equation*}
  The constant $C$ depends only on $r^{-1}$, $a$, $\|\Psi_M\|_\infty$,
  $\|K^\Sigma\|_\infty$, $\|Q\|_\infty$, $\|\RiemL^\Sigma\|_\infty$,
  $\|S\|_\infty$, $\|\RiemM\|_\infty$, and $\inj(M,g)^{-1}$.

  Alternatively, the dependence on $\|\Psi_M\|_\infty$ can be replaced
  by $\|\tilde\Psi_M\|_\infty$.
\end{theorem}
\begin{proof}
  We will restrict to the proof of the first statement, since the
  others are proved similarly.

  We will proceed in a Stampacchia iteration. Let $\eta$ be a cut-off
  function with $\eta\equiv1$ on $B(x,r/16)$ and $\eta\equiv 0$
  outside $B(x,r/8)$ such that $0\leq\eta\leq 1$ and $|\nabla\eta|\leq
  16$. Let $u:= |\chi|$ and for $k\geq 0$ set $u_k:=\max \{\eta u-k, 0\}$.
  In addition set $A(k):= \supp\,u_k$. Then clearly $A(k)\subset
  B(x,r/8)$, such that on $A(k)$ the estimates from theorem
  \ref{thm:sigma_lp} hold.
  
  The $L^2$-bound for $|\chi|$ from lemma \ref{lemma:sigma_l2} implies that
  \begin{equation*}
    k^2 |A(k)| \leq \IAk \eta^2u^2\dmu \leq \int_{B(x,r/8)} \eta^2 u^2 \dmu \leq
    C(r^{-1},a,\|\Psi_M\|_\infty,\|K^\Sigma\|_\infty)\,.
  \end{equation*}
  Therefore there exists
  $k_0=k_0(|\Sigma|,\|\Psi_M\|_\infty,\|K^\Sigma\|_\infty,c_0)<\infty$,
  such that $|A(k)|\leq c_0$ for all $k\geq k_0$. Here we want $c^S_0$
  to be the constant from lemma \ref{lemma:hs_sobolev}, to be able to
  apply the estimate from there for all functions with support in
  $A(k)$, with $k\geq k_0$.
  
  To proceed, let $q>2$.  Multiply the Simons identity, in the form
  \eqref{eq:simons_final} from the proof of proposition
  \ref{prop:sigma_grad}, by $u_k^q$ and integrate. This yields
  \begin{equation*}
    \begin{split}
      &\IAk - u_k^q u \Delta u  + u_k^q(|\nabla\chi|^2 - |\nabla u|^2) \dmu
      \\
      &\quad\leq 
      c \IAk
      u_k^q u^4 
      + |K|u_k^q u^3 + |\RiemL^\Sigma| u_k^q u^2  
      + u_k^q \chi * \nabla \big(Q + \chi * S\big) \dmu\,.
    \end{split}
  \end{equation*}
  Here $c$ is a purely numerical constant. Partially integrate
  the Laplacian on the right hand side and the last term on the left
  hand side. This yields
  \begin{equation*}
    \begin{split}
      &\IAk q \eta u u^{q-1}_k |\nabla u|^2 +
      qu^2u^{q-1}\la\nabla\eta,\nabla u\ra + u_k^q |\nabla\chi|^2 \dmu
      \\
      &\ \leq 
      c\!\!\IAk\!\!
      u_k^q u^4 
      + |K|u_k^q u^3 + |\RiemL^\Sigma| u_k^q u^2  
      + \big(u_k^q|\nabla\chi|  + 
      u_k^{q-1} u|\nabla u_k|\big) \big(|Q| + u|S|\big) \dmu.
    \end{split}
  \end{equation*}
  Note that the term $\int q \eta u u_k^{q-1} |\nabla u|^2 \dmu$ on the
  left hand side controls $\int u_k^q |\nabla u|^2\dmu$. But before we
  use this estimate, we absorb the gradient terms into the first term
  on the left hand side using $|\nabla u_k| \leq \eta\nabla u +
  C(r^{-1}) u$. Consider for example the term containing $|\nabla\chi|^2$:
  \begin{equation*}
    c \IAk u_k^q |\nabla \chi|(|Q|+u|S|) \dmu 
    \leq 
    \IAk u_k^q|\nabla\chi|^2 \dmu + c
    \IAk u_k^q |Q|^2 + u_k^q u^2|S|^2\dmu\,.
  \end{equation*}
  The other terms which contains $|\nabla u|$ can be treated
  similarly, such that the resulting terms can be absorbed on the
  left. This yields an estimate of the form
  \begin{equation}
    \label{eq:stamp}
    \IAk u_k^q |\nabla u|^2 \dmu 
    \leq
    C(q,\|K^\Sigma\|_\infty,\|Q\|_\infty,\|\RiemL^\Sigma\|_\infty,\|S\|_\infty) 
    \IAk u_k^{q-2} u^6 + u_k^{q-2}u^2 \dmu\,.
  \end{equation}
  Note that we used that $u_k\leq u$ and $u\leq u^2 +1$ here to get
  rid of the extra terms. We begin estimating the
  terms on the right hand side of \eqref{eq:stamp} using lemma
  \ref{lemma:hs_sobolev}. Rewrite and estimate the first term as
  follows:
  \begin{equation}
    \label{eq:stamp1}
    \begin{split}
      \IAk u_k^{q-2} u^6 \dmu 
      &= 
      \IAk (u_k u^{6/q-2})^{q-2} \dmu 
      \\
      &\leq 
      |A(k)| \Big( \tilde c_{q-2}^S\IAk\big| \nabla
      (u_k u^{6/q-2})\big|^2 + |Hu_ku^{6/q-2}|^2 \dmu\Big)^{q-2/2}.
    \end{split}
  \end{equation}
  To estimate the first term on the right hand side compute on
  $A(k)$, using $u_k/u\leq 1$,
  \begin{equation*}   
    \big| \nabla (u_k u^{6/q-2})\big| \leq u^{6/q-2}|\nabla u_k| + \tfrac{6}{q-2}
    u^{6/q-2} |\nabla u| \tfrac{u_k}{u} \leq c(q,r^{-1}) (u^{6/q-2}|\nabla u|
    + u^{\frac{6}{q-2}+1}).
  \end{equation*}
  Observe, that if $q$ is large enough, namely such that $2+\tfrac{12}{q-2} <
  p_0$ and $\frac{12}{q-2}+2 \leq 2+p_0$, then theorem \ref{thm:sigma_lp} yields that
  \begin{equation*}
    \IAk \big|\nabla (u_k u^{6/q-2})\big|^2 \dmu\leq C(q) \IAk
    u^{12/q-2}|\nabla u|^2 + u^{\frac{12}{q-2}+2} \dmu \leq C(q).
  \end{equation*}
  Here, and for the remainder of the proof,  $C(q)$ denotes a constant
  that depends on $q$ and, in addition to that, on all the quantities
  the constant in the statement of this theorem depends on. 

  To address the second term in \eqref{eq:stamp1}, recall that
  since $0=\theta^+=H+P$, we have $\|H\|_\infty = \|P\|_\infty \leq
  2 \|K^\Sigma\|_\infty$. Therefore 
  \begin{equation*}
    \IAk H^2 u_k^2 u^{12/q-2} \dmu \leq 4 \|K^\Sigma\|_\infty^2
    \int_\Sigma u^{\frac{12}{q-2}+2}\dmu \leq C(q)\,,
  \end{equation*}
  where the last estimate also follows from theorem \ref{thm:sigma_lp}
  if $q$ is large enough. Summarizing these steps, we have
  \begin{equation*} 
    \IAk u_k^{q-2} u^6 \dmu \leq C(q) |A(k)|\,.
  \end{equation*}
  A similar procedure for the remaining terms
  in \eqref{eq:stamp} finally yields the estimate
  \begin{equation}
    \label{eq:stamp2}
    \IAk u_k^q |\nabla u|^2 \dmu 
    \leq 
    C(q)|A(k)|\,,
  \end{equation}
  provided $q>q_0$ is large enough. Fix such a $q>q_0$ and let $f =
  u_k^{1 + q/2}$. Since
  \begin{equation*}
    |\nabla f|^2 \leq C(q,r^{-1})(u_k^q|\nabla u|^2 + u_k^q u^2),
  \end{equation*}
  equation \eqref{eq:stamp2} and the above estimates imply that
  \begin{equation*}
    \int_{A((k)} |\nabla f|^2 \dmu \leq C(q) |A(k)|.
  \end{equation*}
  The Hoffman-Spruck-Sobolev inequality from lemma
  \ref{lemma:hs_sobolev}, combined with theorem \ref{thm:sigma_lp}, furthermore yields
  \begin{equation*}
    \IAk f^2\dmu = \IAk\!\! u_k^{q+2} \dmu \leq C(q) |A(k)|
    \left(\IAk\!\! |\nabla u|^2 +H u^2\dmu \right)^{\frac{q+2}{2}}
    \leq C(q) |A(k)|\,.
  \end{equation*}
  Thus one further application of lemma \ref{lemma:hs_sobolev} yields
  \begin{equation*}
    \IAk u_k^{q +2} \dmu = \IAk f^2 \dmu \leq C(q)
    |A(k)|^2\,.
  \end{equation*}
  Consider $h>k\geq k_0$, then on $A(h)$ we have that $u_k \geq h-k$ and
  therefore we derive the following iteration inequality
  \begin{equation*}
    |h-k|^{q+2}|A(h)|\leq \IAh u_k^{q+2}\dmu \leq \IAk
     u_k^{q+2}\dmu \leq C(q) |A(k)|^2\,.
  \end{equation*}
  The lemma of Stampacchia \cite[Lemma 4.1]{Stampacchia:1966} now
  implies that $|A(k_0+d)| = 0 $ for
  \begin{equation*}
    d^{q+2} \leq C(q) |A(k_0)| \leq C(q) a
  \end{equation*}
  In view of the definition of $A(k) = \supp \max\{\eta u-k,0\}$, this
  yields in particular $|\chi|(x)|\leq C(q)$. As $x$ was arbitrary,
  the claim follows.
\end{proof}
\begin{corollary}
  \label{cor:main}
  Let $\Sigma\subset M$ be an $L_M$-stable MOTS with $(r,a)$-locally
  uniformly finite area. Then $\Sigma$ satisfies the following
  estimates.
  \begin{equation*}
    \sup_\Sigma |\chi| \leq C(a, r^{-1},\|K\|_\infty,\|\nabla
    K\|_\infty, \|\RiemM\|_\infty, (\mathrm{inj}(M,g))^{-1})
  \end{equation*}
  In addition, for all $x\in\Sigma$ we have an $L^2$-gradient estimate
  of the form
  \begin{equation*}
    \int_{B(x,r/8)} |\nabla\chi|^2\dmu \leq C(a, r^{-1},\|K\|_\infty,\|\nabla
    K\|_\infty, \|\RiemM\|_\infty)\,.
  \end{equation*}
\end{corollary}
\begin{proof}
  The statement to prove is that the constants only depend on the
  stated quantities. This is due to the following reasons.

  First, we use the above estimates using the representation of
  constants containing $\|\tilde\Psi_M\|_\infty$ an
  $\|\RiemM\|_\infty$. As we have seen in remark
  \ref{rem:rewrite_psi}, we can estimate
  \begin{equation*}
    |\tilde\Psi_M| \leq c (|K|^2 + |\nabla K| +  |\RiemM|),
  \end{equation*}
  where $c$ is a numerical constant. Second, since for all $X,Y,Z\in\CX(\Sigma)$
  \begin{equation*}
    Q(X,Y,Z) = \RiemL(X,Y,n,Z) + \RiemL(X,Y,\nu,Z),
  \end{equation*}
  we can use the Gauss and Codazzi equations of the embedding
  $M\hookrightarrow L$ to estimate
  \begin{equation*}
    |Q| + |\RiemL^\Sigma| \leq c(|K|^2 + |\nabla K| + |\RiemM|).
  \end{equation*}  
  Third, obviously 
  \begin{equation*}
    |K^\Sigma|^2 + |S|^2 \leq |K|^2\,.
  \end{equation*}
  Thus we see that all quantities are controlled by $\|K\|_\infty$,
  $\|\nabla K\|_\infty$ and $\|\RiemM\|_\infty$, where the
  $\infty$-norms are computed on $\Sigma$. Note that the
  dependency on $\mathrm{inj}(M)$ comes from the fact that the
  constants $c_0^S$ and $c_1^S$ in the Hoffman-Spruck-inequality only
  depend on $\|\RiemM\|_\infty$ and $\mathrm{inj}(M)^{-1}$. 
\end{proof}
We conclude with an estimate for the principal eigenfunction to $L_M$
or $L_-$.
\begin{theorem}
  \label{thm:eigenfct}
  Let $\Sigma$ be an $L_M$-stable MOTS.  Let $\lambda\geq 0$ be the
  principal eigenvalue of $L_M$ and $f>0$ its corresponding
  eigenfunction. They satisfy the estimates
  {%
    \smallmathindent
    \begin{align*}
      &\lambda|\Sigma| + \tfrac{1}{2} \int_\Sigma f^{-2} |\nabla f|^2 \dmu
      \leq 
      4\pi + \int_\Sigma |S|^2\dmu - \int_\Sigma \tilde\Psi_M \dmu
      \intertext{and}     
      &\begin{aligned}
        \int |\nabla^2 f|^2 \dmu
        &\leq
        C(|\Sigma|, \|K\|_\infty, \|\nabla K\|_\infty,
        \|\RiemM\|_\infty, \inj(M,g)^{-1})
        \!\int_\Sigma f^2 + |\nabla f|^2\dmu
        \\
        &\quad + \lambda^2\!\int_\Sigma f^2\dmu\,.
      \end{aligned}
    \end{align*}
  }%
  The same estimates hold for $L_-$-stable MOTS when $f$ and $\lambda$
  are the principal eigenfunction and eigenvalue of $L_-$ instead,
  then $\tilde \Psi_M$ has to be replaced by $\tilde\Psi_-$ in the first estimate.
\end{theorem}
\begin{proof}
  The first estimate follows from a computation similar to the proof
  of lemma \ref{lemma:LM_sym_stab}.
  
  The second estimate then follows from the first by using the
  identity
  \begin{equation*}
    \int_\Sigma |\nabla^2 f|^2 \dmu = \int_\Sigma (\Delta f)^2 +
    \RicSig(\nabla f, \nabla f)\dmu \,.
  \end{equation*}
  To estimate the terms on the right hand side, note that
  \begin{equation*}
    -\Delta f 
    =
    \lambda f - 2S(\nabla f) - f(\div S - \tfrac{1}{2}|\chi|^2 -|S|^2
    + \tfrac{1}{2}\ScalSig - \tilde\Psi_M)
  \end{equation*}
  and as $\Sigma$ is two-dimensional
  \begin{equation*}
    \RicSig(\nabla f, \nabla f) = \tfrac{1}{2}\ScalSig|\nabla f|^2 \,.
  \end{equation*}
  In view of the Gauss equation for $\Sigma\subset M$ and
  the bounds for $\chi$, we find the claimed estimate.
\end{proof}
\begin{corollary}
  \label{coro:eigenfct}
  If $\Sigma$ is an $L_M$-stable MOTS, then the principal
  eigenfunction $f>0$ to $L_M$ which is normalized such that $\|f\|_\infty=1$
  satisfies the estimate
  \begin{equation*}
    \int_\Sigma f^2 + |\nabla f|^2 + |\nabla^2 f|^2 \dmu 
    \leq
    C(|\Sigma|, |\Sigma|^{-1},\|K\|_\infty, \|\nabla K\|_\infty,
    \|\RiemM\|_\infty, \inj(M,g)^{-1})
  \end{equation*}
  The same estimate holds for $L_-$-stable MOTS, when $f$ is the
  principal eigenfunction to $L_-$ instead, and the constant depends
  on  $|\Sigma|$, $|\Sigma|^{-1}$, $\|K\|_\infty$,
  $\|\RiemL\|_\infty$, and $\inj(M,g)^{-1})$.
\end{corollary}
\begin{proof}
  Since $\|f\|_\infty=1$, we have $\int_\Sigma f^2\dmu
  \leq |\Sigma|$. Then since $f^{-2}\geq 1$, the first estimate from
  the previous theorem implies
  \begin{equation*}
    \int_\Sigma |\nabla f|^2 \leq C(|\Sigma|,\|K\|_\infty, \|\nabla
    K\|_\infty, \|\RiemM\|_\infty)\,.
  \end{equation*}
  Since $ \lambda^2 \int_\Sigma f^2 \leq \lambda^2 |\Sigma| \leq C|\Sigma|^{-1}$
  the above estimates combined with the previous theorem imply the
  claim.
\end{proof}
\begin{remark}
  A local version of these estimates can also be derived from local
  area bounds, like the curvature estimates before. In the subsequent
  application, however, we will not use this more general form.
\end{remark}

%% file: areabound.tex
\section{Local area bounds}
\label{sec:area}
This section is devoted to derive the area bounds needed for the
curvature estimates in the previous section.

The following theorem is analogous to Pogorelov's estimate for stable
minimal surfaces \cite{Pogorelov:1981}. We will modify the proof of
Colding and Minicozzi given in \cite{Colding-Minicozzi:2002}.
\begin{theorem}
  \label{thm:area_bound}
  Let $(M,g)$ be a Riemannian 3-manifold with curvature bounded above
  $\|\RiemM\|_\infty\leq C$.

  Let $\Sigma\subset M$ be an immersed surface with bounded mean
  curvature $\|H\|_\infty\leq C$. Assume that there exist $\alpha>0$
  and a constant $Z\geq 0$ such that for all functions $\eta\in C^\infty_c(\Sigma)$ we have
  \begin{equation}
    \label{eq:stab}
    -\int_\Sigma \ScalSig\,\eta^2 \dmu
    \leq
    \int_\Sigma (2-\alpha) |\nabla\eta|^2 + Z\eta^2 \dmu.
  \end{equation}
  Then there exists $r_0 = r_0(\alpha^{-1}, Z, \|H\|_\infty,\|\RiemM\|_\infty)$
  such that for all $r<r_0$, the area $|\Sigma \cap B_\Sigma(x,r)|$
  is bounded by
  \begin{equation*}
    |\Sigma \cap B_\Sigma(x,r)| \leq \tfrac{4\pi}{\alpha} r^2.
  \end{equation*}  
\end{theorem}
\begin{proof}
  Fix an arbitrary $x\in\Sigma$. By the Gauss equation for $\Sigma$ we know that
  \begin{equation*}
    \ScalSig = \ScalM - 2\RicM(\nu,\nu) + H^2 - |A|^2 \leq C(\|H\|_\infty,\|\RiemM\|_\infty).
  \end{equation*}
  Hence by the Rauch comparison theorems (cf. \cite[Section
  1.10]{Cheeger-Ebin:1975}), there is a radius
  $0<r_1=r_1(\|H\|_\infty,\|\RiemM\|_\infty)$ such that $\Sigma$ has
  no conjugate points in $B_\Sigma(x,r_1)$. Hence the pull-back
  $\gamma$ of the metric of $\Sigma$ to the disc $D_{r_1}:=B(0,r_1)$
  in $T_x\Sigma$ is regular and satisfies \eqref{eq:stab}. 

  Denote $D_s = B(0,s)$ the disk of radius $0\leq s \leq r_1$ in
  $D_{r_1}$ and $\Gamma_s=\del D_s$ the boundary. Note that $D_s$ is a
  topological disk and $\Gamma_s$ is a single circle. Furthermore the
  area of $D_s$ with respect to $\gamma$ is bigger than
  $|B_\Sigma(x,s)|$, for $s<r_1$.

  In the stability inequality, we set $\eta=\eta(s)=
  \max\{1-\tfrac{s}{r_0}, 0\}$, where $0<r\leq r_1$ will be chosen
  below. Denote by $K(s) = \int_{D_s} \Scal \dmu$. Then
  $K'(s)=\int_{\Gamma_s} \Scal \dl$, where $\dl$ is the line element
  of $\Gamma_s$ induced by $\gamma$. Hence, by the co-area formula and
  partial integration
  \begin{equation*}
    - \int_{D_r} \Scal\eta^2\dmu = - \int_0^r K'(s) \eta^2(s) \ds =
    \int_0^r K(s) (\eta^2(s))'\ds.
  \end{equation*}
  Let $l(s)=\length(\Gamma_s)$. By the formula for the variation of
  arc length and the Gauss-Bonnet formula we find $l'(s) = 2\pi -
  K(s)$, as $\Gamma_s$ is a circle, on which the geodesic curvature
  integrates to $2\pi$. Thus we compute, using
  $(\eta^2(s))'=-\frac{2}{r}(1-\frac{s}{r})$, $(\eta^2(s))'' =
  \frac{2}{r^2}$ and the co-area formula
  \begin{equation*}
    \begin{split}
      - \int_{D_r} \Scal\eta^2\dmu
      &=
      2\pi \int_0^r (\eta^2(s))'\ds + \int_0^r l(s)(\eta^2(s))''\ds
      \\
      &=
      -\frac{2\pi}{r}\int_0^r\big(1-\frac{s}{r}\big)\ds + \frac{2}{r^2} \int_0^r l(s) \ds
      = - 2\pi + \frac{2}{r^2} |D_r|
    \end{split}
  \end{equation*}
  Furthermore, compute
  \begin{equation*}
    \int_{D_r} |\nabla\eta|^2\dmu = \frac{1}{r^2}\int_0^r l(s)\ds = \frac{1}{r^2}|D_r|,
  \end{equation*}
  and estimate
  \begin{equation*}
    \int_{D_r} Z\eta^2 \dmu \leq Z |D_r|.
  \end{equation*}
  Hence equation \eqref{eq:stab} implies that
  \begin{equation*}
    -2\pi + \tfrac{2}{r^2}\,|D_r| \leq \tfrac{2-\alpha}{r^2}\, |D_r| + Z |D_r|.
  \end{equation*}
  Thus
  \begin{equation*}
    \alpha |D_r| \leq 2\pi r^2 + Z r^2 |D_r|.
  \end{equation*}
  Choose $r^2 = \min\{\tfrac{\alpha}{2Z}, r_1^2\}$ and absorb the error
  term. This yields the claim.
\end{proof}
The above theorem yields the local area bound for stable MOTS needed
for the curvature estimates.
\begin{corollary}
  \label{coro:area_bound_MOTS}
  Let $\Sigma$ be an $L_M$-stable MOTS. Then there exists $r_0>0$
  depending only on $\|K\|_\infty$, $\|\nabla K\|_\infty$, and
  $\|\RicM\|_\infty$, such that for every $x\in\Sigma$ and $r<r_0$
  \begin{equation*}
    |\Sigma\cap B_\Sigma(x,r)| \leq 6\pi r^2.
  \end{equation*}
\end{corollary}
\begin{proof}
  To see that theorem \ref{thm:area_bound} is applicable on an
  $L_M$-stable MOTS, recall equation~\eqref{eq:LM_sym_stab_GS}.
  As $|\chi|^2 \geq \frac{1}{2} |A|^2 - 4|K^\Sigma|^2$ and by the
  Gauss equation
  \begin{equation*}
    |A|^2 = \ScalM-\ScalSig-2\RicM(\nu,\nu) + H^2,
  \end{equation*}
  we find from \eqref{eq:LM_sym_stab_GS}, taking the scalar curvature
  term to the left, that
  \begin{equation*}
    -\frac{3}{2} \int_\Sigma \ScalSig \eta^2\dmu \leq \int_\Sigma
    2|\nabla\eta|^2 + Z\eta^2 \dmu
  \end{equation*}
  where $Z = Z(\|K\|_\infty,\|\nabla K\|_\infty, \|\RicM\|_\infty)$.
  As on a MOTS $\|H\|_\infty = \|P\|_\infty \leq
  2\|K^\Sigma\|_\infty$, theorem \ref{thm:area_bound} gives the
  desired bounds.
\end{proof}
Theorem \ref{thm:main} and corollary \ref{cor:main} imply the
following estimates.
\begin{corollary}
  \label{coro:curvature_bound_MOTS}
  Let $\Sigma$ be an $L_M$-stable MOTS, then
  \begin{equation*}
    \|\chi\|_\infty \leq C\big(\|K\|_\infty,\|\nabla
    K\|_\infty,\|\RiemM\|_\infty, (\inj(M,g))^{-1}\big).
  \end{equation*}
  Furthermore, there exists $0<\bar r=\bar r(\|K\|_\infty,\|\nabla
    K\|_\infty,\|\RiemM\|_\infty)$ such that for all
  $x\in\Sigma$
  \begin{equation*}
    \int_{B(x,\bar r)} |\nabla\chi|^2\dmu \leq C(\|K\|_\infty,\|\nabla
    K\|_\infty, \|\RiemM\|_\infty)\,.
  \end{equation*}
\end{corollary}


%% file: applic.tex
\section{Applications}
\label{sec:applic}
The main application of the curvature estimates proved in this paper
is the following compactness property of stable MOTS.
\begin{theorem}
  \label{thm:compact}
  Let $(g_n, K_n)$ be a sequence of initial data sets on a manifold
  $M$. Let $(g,K)$ be another initial dataset on $M$
  such that
  \begin{align*}
    \|\RiemM\|_{\infty}&\leq C\,,
    \\[.2ex]
    \|K\|_{\infty} + \|\nabM K\|_{\infty} &\leq C, 
    \\[.2ex]
    \mathrm{inj} (M,g) &\geq C^{-1}\,,
    \intertext{for some constant $C$. Assume that}
    g_n \to g \quad&\text{in}\quad C_\text{loc}^2(M,g)\ \text{and},\\
    K_n \to K \quad&\text{in}\quad C_\text{loc}^1(M,g).
  \end{align*}
  Furthermore, let $\Sigma_n\subset M$ be a sequence of immersed
  surfaces which are stable marginally outer trapped with respect to
  $(g_n, K_n)$ and have an accumulation point in $M$. In addition,
  assume that the $\Sigma_n$ have uniformly locally finite area, that
  is, for all $x\in M$ there exists $0<r=r(x)$ and $a=a(x)<\infty$
  such that
  \begin{equation}
    \label{eq:app_ab}
    |\Sigma_n \cap B_{M_{t_n}}(x,r)| \leq ar^2 \qquad
    \text{uniformly in\ } n,
  \end{equation}
  where $B_{M_{t_n}}(x,r)$ denotes the ball in $M$ around $x$ with
  radius $r$.

  Then a subsequence of the $\Sigma_n$ converges to a smooth immersed
  surface $\Sigma$ locally in the sense of $C^{1,\alpha}$ graphs.
  $\Sigma$ is a MOTS with respect to $(g, K)$. If $\Sigma$ is
  compact, then it is also stable.  \fillbox
\end{theorem}
\begin{proof}
  By the estimates in corollary \ref{cor:main}, the above assumptions,
  even without \eqref{eq:app_ab}, are sufficient to imply that the
  shears $\chi_n$, and thus the second fundamental forms $A_n$, of the
  $\Sigma_n$, with respect to the metric $g_n$ are uniformly bounded
  \begin{equation*}
    |A_n| \leq C.
  \end{equation*}
  As $g_n$ is eventually $C^2$-close to $g$, this bound translates to
  a bound for the second fundamental forms $\tilde A_n$ of $\Sigma_n$
  with respect to the metric $g$.
  
  In the sequel $B_M(x,s)$ denotes an extrinsic ball in $(M,g)$. As
  the geometry of the $(M,g_n)$ is uniformly bounded, the uniform
  curvature bound implies in particular, that there exists a radius
  $s$ such that for every $x\in\Sigma_n$, the connected component of
  $\Sigma_n \cap B_{M}(x,s)$ containing $x$ can be written as graph of
  a function $u_n^x$ over $T_x\Sigma_n$, where the function $u_n^x$ is
  uniformly bounded in $C^2$. Without loss of generality, we can
  assume that $s<r$, where $r$ is from equation \eqref{eq:app_ab}.

  Now let $x\in M$ be arbitrary. From the previous fact we conclude
  that each connected component of $\Sigma_n \cap B_{M}(x,s)$,
  which intersects $\Sigma_n \cap B_{M}(x,s/2)$ contains a
  uniform amount of area. In view of the local area bound
  \eqref{eq:app_ab}, we conclude that there are only finitely many
  such components. Furthermore, the maximal number of those components
  is uniform in $n$.

  Hence, in each ball $B_{M}(x,s/2)$, we can extract a convergent
  subsequence of the $\Sigma_n$, such that $\Sigma_n \cap B_{M}(x,s)$
  converges in $C^{1,\alpha}$ to a smooth surface $\Sigma^x$.
  As the $M_{t_n}$ can be covered by countably many such balls, a
  diagonal argument yields a convergent subsequence of the $\Sigma_n$
  and a limit surface $\Sigma$, which is immersed (cf. remark
  \ref{rem:applic}). Note that since the $\Sigma_n$ have an
  accumulation point in $(M,g)$, the limit $\Sigma$ is
  non-empty.

  Furthermore, $C^{1,\alpha}$ convergence yields that $\Sigma$
  is $C^{1,\alpha}$ and satisfies a weak version of the equation
  $\theta^+=0$. In view of standard regularity theory for prescribed
  mean curvature equations, we find that $\Sigma$ is in fact
  smooth (cf. \cite{Gilbarg-Trudinger:1998}).
  
  If $\Sigma$ is compact, we can cover $\Sigma$ with
  finitely many balls $B(x,s/8)$. As before we know that locally the 
  $\Sigma_n$ converge to $\Sigma$ in $C^{1,\alpha}$. Since we also
  have local $W^{1,2}$-bounds on $\chi$ we can furthermore assume that
  $\Sigma_n\to\Sigma$ in $W^{2,p}$ for a fixed, large $p$, which
  will be selected below.
  
  From this we can conclude that, the metrics of the $\Sigma_n$
  converge to the metric of $\Sigma$ in $C^\alpha\cap W^{1,p}$.
  We can pull back the metrics of the $\Sigma_n$ to $\Sigma$
  and call them $\gamma_n$.  The metric on $\Sigma$ will be
  denoted by $\gamma$. Then define the operators $L_n$ as the pull
  backs of the operator $L_M$ on $\Sigma_n$ to $\Sigma$. Let $f_n$ be
  the principal eigenfunctions of $L_n$ with eigenvalues $\lambda_n$
  and normalize such that $\|f_n\|_\infty=1$. Since the area of the
  $\Sigma_n$ is eventually bounded below by half of the area of
  $\Sigma$, theorem \ref{thm:eigenfct} implies that $0\leq
  \lambda_n\leq C$, where $C=C(\bar C, \|K\|_\infty,\|\nabla
  K\|_\infty,\|\RiemM\|_\infty)$.  Thus we can assume that the
  $\lambda_n$ converge to some $\lambda$ with $0\leq \lambda \leq C$.
  
  By corollary \ref{coro:eigenfct}, the $W^{2,2}$-norm of the $f_n$
  taken with respect to the metrics $\gamma_n$ is uniformly bounded.
  Recall that the difference of the Hessian of $f$ with respect to
  $\gamma^n$ and $\gamma$ is of the form
  \begin{equation*}
    \big(\nabla^2_{\gamma_n} - \nabla^2_{\gamma} \big) f =
    -\big(\Gamma_{\gamma_n} - \Gamma_{\gamma}\big) * df
  \end{equation*}
  where $\Gamma_\gamma$ and $\Gamma_{\gamma_n}$ denote the connection
  coefficients of $\gamma$ and $\gamma_n$. Furthermore $\nabla f$ is bounded
  in any $L^p$ and by $W^{1,p}$ convergence of the metrics
  $\Gamma_{\gamma_n} - \Gamma_{\gamma}\to 0$ in $L^p$. Thus we find that also
  $\|f_n\|_{W^{2,2}}\leq C$, where the norm is taken with respect to
  the metric $\gamma$ on $\Sigma$.
  Hence we can assume that $f_n\to f$
  in $W^{1,p}$. The Sobolev embedding $W^{1,p}\hookrightarrow C^0$,
  implies that $f\geq 0$, and $\|f\|_\infty=1$, so $f\not\equiv 0$.

  The next step is to take the equation $L_n f_n =\lambda_n f_n$ to the
  limit. Since $f_n\to f$  only in $W^{1,p}$, we have to use the weak
  version of this equation, namely that for all $\phi\in
  C^{\infty}(\Sigma)$ 
  \begin{equation*}
    \int_{\Sigma} \gamma_n^{ij} (d f_n)_i d \phi_j +  B_n^i (d f_n)_i\phi
    + C_n f\phi \dmu = \lambda_n \int_{\Sigma} f_n \phi \dmu\,,
  \end{equation*}
  where $B_n$ and $C_n$ are the coefficients of the operator
  $L_n$. By the $W^{2,p}$-convergence of the surfaces, we find that
  $\gamma_n$ converges to $\gamma$ in $W^{1,p}$, and $B^i_n$ and $C_n$
  converge in $L^p$ to the coefficients $B^i$ and $C$ of $L_M$ on
  $\Sigma$.
  Thus, since $f_n$ converges in $W^{1,p}$ to $f$, we can choose $p$
  large enough to infer that the limit of the above integrals
  converges to the corresponding integral on $\Sigma$, that is $f$
  satisfies
  \begin{equation*}
    \int_{\Sigma} \la \nabla f, \nabla \phi \ra + \la B, \nabla f
    \ra \phi + C f \phi \dmu = \lambda \int_{\Sigma} f\phi \dmu\,.
  \end{equation*}
  Thus $f$ is a weak eigenfunction of $L_M$ on $\Sigma$. Elliptic
  regularity implies that $f$ is smooth and satisfies $L_M f =
  \lambda f$. Since $\lambda\geq 0$ and $f\geq 0$, $f\not\equiv 0$,
  we conclude that $\Sigma$ is stable.
\end{proof}
\begin{remark}
  \label{rem:applic}
  If the limit surface is not compact, then it still follows that it
  is ``symmetrized'' stable in the sense that inequality \eqref{eq:LM_sym_stab_GS}
  holds for all test functions $\eta$ with compact support.
\end{remark}
\begin{remark}
  \label{rem:applic2}
  \begin{mynum}
  \item  
    If the surfaces $\Sigma_n$ are embedded, one would assume that the
    limit $\Sigma$ is embedded as well. However this is not
    necessarily the case. This is due to the fact that at a point $p$
    where $\Sigma$ touches itself, the equation $\theta^+=0$ is
    satisfied with respect to the outward normal. At $p$ the normals
    corresponding to these two sheets point into opposite directions at
    $p$. If we flip one of the normals, to make them point into the same
    direction to apply the maximum principle in a graphical situation,
    the equation for one sheet will remain $\theta^+=0$, but for the
    other it will change to $\theta^-=0$. Hence, one cannot compare the
    two sheets.
    
    However, if $\Sigma$ also satisfies $\theta^-\leq 0$, then
    the maximum principle implies that the set $S$ of touching points is
    open. By continuity $S$ is closed, and $S\neq \Sigma$, as the
    $\Sigma_n$ are embedded and have bounded curvature. Hence
    $S=\emptyset$ and $\Sigma$ is embedded.
  \item If the assumption of uniformly locally finite area does not
    hold, but the surfaces $\Sigma_n$ are embedded, the limit still
    exists in the sense of laminations. Here, the limit is a
    lamination, for which the leaves are ``symmetrized'' stable MOTS
    (cf. remark \ref{rem:applic}). Convergence is in the sense of
    laminations in the class $C^\alpha$, with convergence of the
    leaves in $C^{1,\alpha}$, for any $0 <\alpha<1$. For a proof of
    this statement, we refer to \cite[Appendix
    B]{Colding-MinicozziIV:2004}, in the reference this is stated for
    minimal laminations, but the modification to the MOTS case is
    straightforward.
    
    For the sake of completeness we state the definition of a
    \emph{lamination}. A \emph{lamination} $\CL\subset M$ is a closed
    set, which is a disjoint union of complete connected smooth
    surfaces, called \emph{leaves}. Furthermore there are coordinate
    charts for $M$, $\psi: V\subset M \to \IR^3$, $V$ a neighborhood of
    some point $x\in M$, such that the image of each leaf $L$ of $\CL$
    is contained in a set of the form $\IR^2\times{t}$, where $t\in I$,
    and $I$ is a closed subset of $\IR$.
    A sequence of laminations $\CL_n$ is said to converge to a
    lamination $\CL$ if the coordinate charts converge and $\CL$ is the
    set of accumulation points of the~$\CL_n$.
  \item There are examples of compact three dimensional manifolds
    which contain sequences of compact stable minimal surfaces of
    fixed genus and unbounded area \cite{Colding-Minicozzi:1999},
    \cite{Dean:2003}. Thus assumption \eqref{eq:app_ab} does not
    follow immediately from standard theory, even when the surfaces
    $\Sigma_n$ are confined to a compact region.
  \end{mynum}
\end{remark}


%% file: ack.tex
\section*{Acknowledgements}
The authors wish to thank Walter Simon, Marc Mars, Greg Galloway, Rick
Schoen and Gerhard Huisken for useful conversations. We are grateful
for the hospitality and support of the Isaac Newton Institute, as part
of this work was done during the workshop \emph{Global Problems in
  Mathematical Relativity}.